\documentclass[10pt,twocolumn,prb,aps,showpacs,longbibliography]{revtex4-1}
\usepackage[latin9]{inputenc}
\usepackage{color}
\usepackage{bm}
\usepackage{amsmath}
\usepackage{amssymb}
\usepackage{graphicx}
\usepackage{mathtools}
\usepackage{subfigure}
\usepackage{hyperref}
\usepackage{esint}

\definecolor{grey}{rgb}{.6,.6,.6}

\newcommand{\cd}[1]{c^{\dagger}_{#1}}

\makeatletter

\usepackage{dcolumn}
\usepackage{bm}
\usepackage{slashed}
\usepackage{amsthm}

\makeatother

\begin{document}

\title{Multidimensional dark space and its underlying symmetries: towards dissipation-protected qubits}

\author{Raul A. Santos$^{1*}$, Fernando Iemini$^{2,3}$, Alex Kamenev$^{4,5}$ and Yuval Gefen$^{6}$}

\affiliation{$^{1}$T.C.M. Group, Cavendish Laboratory, University of Cambridge,
J.J. Thomson Avenue, Cambridge, CB3 0HE, United Kingdom}

\affiliation{$^{2}$ Instituto de F\'isica, Universidade Federal Fluminense, 24210-346 Niter\'oi, Brazil}

\affiliation{$^{3}$ Abdus Salam ICTP, Strada Costiera 11, I-34151 Trieste, Italy}

\affiliation{$^{4}$School of Physics and Astronomy, University of Minnesota,
Minneapolis, Minnesota 55455, USA}

\affiliation{$^{5}$William I. Fine Theoretical Physics Institute, University
of Minnesota, Minneapolis, Minnesota 55455, USA}

\affiliation{$^{6}$Department of Condensed Matter Physics, The Weizmann Institute
of Science, Rehovot 76100, Israel}

\date{January 2020}
\maketitle

{\bf  

Quantum systems are always subject to interactions with an environment, typically resulting in decoherence and 
distortion  of quantum correlations.  It has been recently shown  that a controlled interaction with the environment may actually help to  create a state, dubbed as 
{\em ``dark''}, which is immune to decoherence.  To encode quantum information in the dark states, they need to span a space with a dimensionality  larger than one, so different orthogonal states act as a computational basis. We devise a symmetry-based conceptual framework to engineer such degenerate dark spaces (DDS), 
protected from decoherence by the environment. 
We illustrate this construction with a model protocol, inspired by the fractional quantum Hall effect, where the DDS basis  is isomorphic to a set of degenerate Laughlin states. The long-time steady state of our driven-dissipative model exhibits thus all the characteristics of degenerate vacua of a unitary topological system. This approach offers new possibilities for storing, protecting and manipulating quantum information in open systems.}

It is believed that dissipation conspires against coherence of  quantum states, rendering them to be close to a classical ensemble. This belief was recently challenged by approaches 
aimed at incorporating both drive and dissipation to reach a correlated coherent steady state \cite{Griesser2006,Diehl2008,Diehl2010,Roncaglia2010,Yi2012,Berceanu2016,Zhou2017,Devoret2013,Devoret2016,Siddiqi2016,Murch2019}. 
One remarkable example has been the idea of harnessing dissipation to purify non-trivial topological states \cite{Diehl2011,Bardyn2013,Iemini2016}. This is achieved by a 
careful interplay between radiation-induced drive, and coupling to an external bath, that provides a desired relaxation channel. 
A sequence of excitations and relaxations generates, in the long time limit, a non-equilibrium steady state that decouples from the external 
drive creating a decoherence free subspace \cite{Lidar1998, Knill2000} dubbed a {\it dark} state.
This idea opens a way to engineer a rich variety of non-trivial stationary states, going well beyond thermal states of equilibrium systems.

In order to use this approach to design (and ultimately manipulate) qubits, it is necessary to engineer a non-equilibrium steady {\it space},  which is at least two-dimensional\cite{Iemini2016,Devoret2018}. Here we develop a framework to construct driven dissipative schemes with  degenerate dark spaces (DDS). We achieve this goal  by analyzing the role of symmetries in dissipative dynamics. Specifically, we claim that the dimensionality of DDS is given by the period of the projective symmetry representation\cite{Hamermeshbook}, inherent to the system's evolution (which is considered to be Lindbladian\cite{Lindblad1976,Gorini1976}). 
To this end we extend the discussion of Lindbladian symmetries\cite{Buca2012, Albert2014, Zhang2019} to include those that are realized projectively, providing a link between the projective representations and the dimensionality of the DDS density operator.

We illustrate this framework by studying driven dissipative evolution of a correlated one-dimensional (1D) system, 
inspired by Laughlin quantum Hall states with $\nu=1/m$ filling fractions ($m$ is an odd integer) in a quasi-1D strip (the so-called ``thin torus limit"). This evolution is described 
by a Lindbladian master equation that possesses a DDS. The latter is spanned by $m$ orthogonal  vectors, isomorphic to the set of 
many-body Laughlin ground states on the torus. This correlated DDS has an extra advantage of being exactly described by computationally convenient matrix product states (MPS).  We design a systematic protocol, based on adiabatically varied Lindbladians, that maximizes the purity and fidelity (overlap with the dark space) of its ultimate steady states.

As a warm up for the symmetry discussion, let's consider the Hamiltonian case. If a Hamiltonian is invariant under the action of a symmetry group $\mathcal{G}$, then action of the group elements, $g\in \mathcal{G}$,  on a state is 
implemented by a unitary representation\cite{Hamermeshbook}. In particular, the eigenstates of the Hamiltonian can be labeled by  eigenvalues of the symmetry operator.
As quantum states form rays in the Hilbert space, such that states $|\psi\rangle$ and $e^{i\phi}|\psi\rangle$ are equivalent, it is natural to consider representations that 
satisfy the group multiplication rule  up to a phase, i.e. {\em projective representations}\cite{Hamermeshbook}, defined as 
\begin{equation}\label{proj_rep}
 D(g_1)D(g_2)= e^{i\phi(g_1,g_2)}D(g_1g_2),
\end{equation}
where $D(g)$ is a representation of a group element $g\in \mathcal{G}$. 
Every projective representation is characterized by the set of phases $\omega_2(g_1,g_2)=e^{i\phi(g_1,g_2)}$, known as a 2-cocycle, which are strongly constrained by 
associativity\cite{Hamermeshbook}. For a quantum system invariant under the projective representation (\ref{proj_rep}),
the {\em period} of the 2-cocycle (i.e the minimum number $m$ such that $[\omega_2(g_1,g_2)]^m=1$ for all $g_1,g_2$) determines the dimension of the degenerate space.
Given  a non-trivial 2-cocycle,  representations of at least some group elements do not commute, even for an Abelian group $\mathcal{G}$, i.e. for some $g, h\in \mathcal{G}$,   
$[D(g),D(h)]\neq 0$. One can thus label the eigenstates by  eigenvalues of say  $D(g)$ and generate a different state, with the same energy by acting on it with $D(h)$. 
Notice that this argument implies degeneracy of an entire spectrum.

{\bf Degenerate dark states in Lindbladian evolution.-}
The way symmetry operators act on the Lindbladian  operators is rather different from the Hamiltonian case. In a system with a combined unitary and dissipative dynamics, the most general  Markovian evolution of the density matrix operator, 
$\rho$,  is described  
by the quantum master equation,  $\dot{\rho}= \mathcal{L}(\rho)$, with
\begin{equation}\label{proj_rep_eq}
\mathcal{L}(\rho)= -i[H,\rho]+ \sum_i\left(\ell_i\rho \ell_i^\dagger -\frac{1}{2}\left(\ell_i^\dagger \ell_i\rho+\rho\ell_i^\dagger \ell_i\right)\right).
\end{equation}
Here $H$ is an effective Hamiltonian that represents the unitary evolution. 
Generically non-Hermitian, quantum jump operators, $\ell_i$, describe environment-induced dissipation effects\cite{Lindblad1976,Gorini1976,breuer2002book}. 

A Lindbladian is invariant under an irreducible unitary representation $D(g)$ with $g$ an element of $\mathcal{G}$, if the Hamiltonian 
and the quantum jump operators satisfy \cite{Buca2012, Albert2014}
\begin{equation}\label{symmetries}
 D(g) H D^\dagger(g)=H; \quad 
 \quad D(g) \ell_i D^\dagger(g)=\sum_j \mathcal{U}_{ij}^{(g)}\ell_j,
\end{equation}
(also known as {\it weak symmetry} \cite{Buca2012}) where $\mathcal{U}^{(g)}$ is a unitary matrix that depends on $g$. In particular, if $\sigma$ is an eigenmatrix of $\mathcal{L}$ with an eigenvalue $\lambda_\sigma$, i.e. 
$\mathcal{L}(\sigma)=\lambda_\sigma\sigma$, then
$D(g)\sigma D^\dagger(g)$ is an eigenmatrix with the same eigenvalue. These eigenmatrices obtained from
$\sigma$ by conjugation can represent either the same, or different  states. For a projective representation $D(g)$,  satisfying (\ref{proj_rep}),
$D(g)\sigma D^\dagger(g)$ and $\sigma$
are necessarily different for some element $g$. This is because if for a particular element $h\in\mathcal{G}$, $D(h)\sigma D^\dagger(h)= \sigma$, then  $\sigma$ and $D(h)$
share eigenvectors, so one can take an element $g$ such that $D(g)$ does not commute with $D(h)$. Given that $[D(g),D(h)]\neq 0$, 
$D(g)\sigma D^\dagger(g)$ and $\sigma$ do not share eigenvectors, meaning that they are different. By virtue of the Schur's lemma\cite{Hamermeshbook}, the only case where this logic fails is for $\sigma$ a 
fully mixed state, proportional to the identity matrix.

Focusing on the case of projective representations of Abelian groups, where $D(h)D(g)=e^{i\tilde{\phi}(h,g)}D(g)D(h)$ (here $\tilde{\phi}(g_1,g_2)=\phi(g_1,g_2)-\phi(g_2,g_1)$ is again a cocycle)  one can 
determine the dimension of the degenerate subspace in terms of the factor $e^{i\tilde\phi(h,g)}$. Focusing on an eigenvector $|e\rangle$ of 
$D(h)$ with eigenvalue $e^{i\alpha}$, the operator $D(g)$ acts on this state as a cyclic raising operator, as $D(h)D(g)|e\rangle = e^{i(\alpha+\tilde\phi(h,g))}D(g)|e\rangle$.
For $\tilde\phi(h,g)=\frac{2\pi a}{m}$, with $a,m$ co-prime integers, one can raise a state $m$ times before it returns to itself. This implies that {\em all} eigenspaces of $\mathcal{L}$ are $m$-fold degenerate. 
In particular, the stationary subspace, defined by the eigenvalue $\lambda=0$, is $m$-dimensional. 
Note that the degeneracies of the eigenspaces of $\mathcal{L}$ do not translate into degeneracies of the density matrix in general, as the the eigenmatrices of
$\mathcal{L}$ need not to be self-adjoint, but can come in pairs $\{\sigma,\sigma^\dagger\}$ with eigenvalues $\{\lambda_\sigma,\lambda_\sigma^*\}$.
For the DDS, this is not a problem as $\lambda=0$.

Although conditions presented above are sufficient for the existence of DDS, they are actually excessive and impractical.  
Indeed, the symmetry operators $D(g)$ split the entire Hilbert space into sectors of different quantum numbers that do not mix during the evolution. In other words, there is a certain number of conservation laws, 
which confine the long time evolution of any initial state to only a limited fraction of  DDS.   To access the entire DDS, this needs to be avoided.
In order to achieve this, we consider states ${\bm \rho}$ in the DDS that satisfy
\begin{equation}\label{frust_free}
 \ell_i {\bm \rho}=0 \,\,\mbox{for all }i.
 \end{equation}
We call such DDS {\it frustration free}, as ${\bm \rho}$ is annihilated individually by each quantum jump operator, $\ell_i$.
For systems that satisfies Eq.~(\ref{frust_free}), one can deform quantum jump operators in a way that the symmetry is broken in all the decaying subspaces, while it is maintained within the DDS.
With this in mind, we define {\em dressed} quantum jump operators, that do \textit{not}  satisfy Eq.~(\ref{symmetries}), as $\tilde{\ell}_i=R^\dagger_i \ell_i$, where $R^\dagger_i$ are for now arbitrary local operators.  Dynamics, generated by the dressed  operators,  does not,  in general,
obey any conservation laws. Yet, $\tilde{\ell}_i$ still satisfy Eq.~(\ref{frust_free}), which preserves the DDS and its degenerate multidimensional nature.   For properly dressed quantum jump operators, a generic initial state evolves into a state within DDS, which has projections on all of its basis eigenmatrices.  

Below we demonstrate these considerations on a 1D model, borrowing intuition from the well-studied physics of the Laughlin states on a torus \cite{Haldane1985,Hermanns2008}.  
In particular, we demonstrate that the system is driven to DDS regardless of the nature of an initial state (pure or mixed). We also devise adiabatic  time-dependent Lindbladians that guarantee that the initial state is fully driven into the DDS, resulting in a state with a maximized purity.

{\bf Laughlin states in a narrow torus geometry.-}

A quantum Hall droplet of $N$ electrons subject to a magnetic flux $N_\Phi=m N$ (in units of the flux quanta) and filling fraction $N/N_\Phi\equiv\nu=1/m$ (with  odd integer $m$) 
develops nontrivial correlations that are reproduced by  Laughlin wavefunctions\cite{Laughlin1981}. 
In a torus with periods $L_x$ and $L_y$ (with distances measured in units of the magnetic length), the area is related to the flux as $L_xL_y=2\pi N_\Phi$. 
 The Laughlin states at filling $\nu=1/m$ correspond to exact 
zero energy states of a local Hamiltonian\cite{Trugman1985}, which, after projecting onto the lowest Landau level (LLL), takes the form  
$\mathcal{H}=\sum_n \big(\ell^\dagger_{0,n}\ell_{0,n}+\ell^\dagger_{1,n}\ell_{1,n}\big)$, where the operators $\ell_{s,n}$ $(s=0,1)$ are\cite{Ortiz2013} (see supplemental material)
\begin{equation}\label{Q_operator_laughlin}
 \ell_{s,n}=\sum_{l\geq0}\eta\left({l+\frac{s}{2}}\right) c_{n-l-s}c_{n+l}.
\end{equation}
Here $c_n$ destroys an electron at orbital $n$; $\eta(l)\propto e^{-(\kappa l)^2}$ is a fast decaying function in the narrow torus limit, $\kappa^2=\frac{2\pi}{N_\Phi}\frac{L_x}{L_y}\gg1$,  
(see supplemental material).  The crucial property of Laughlin states $|\Psi\rangle$, that makes them useful for our discussion of the Lindbladian evolution,  is that they  satisfies $\ell_{s,n}|\Psi\rangle=0$, for $s=0,1$ and all $n$.

\begin{figure}[ht!]
\includegraphics[width=0.8\linewidth]{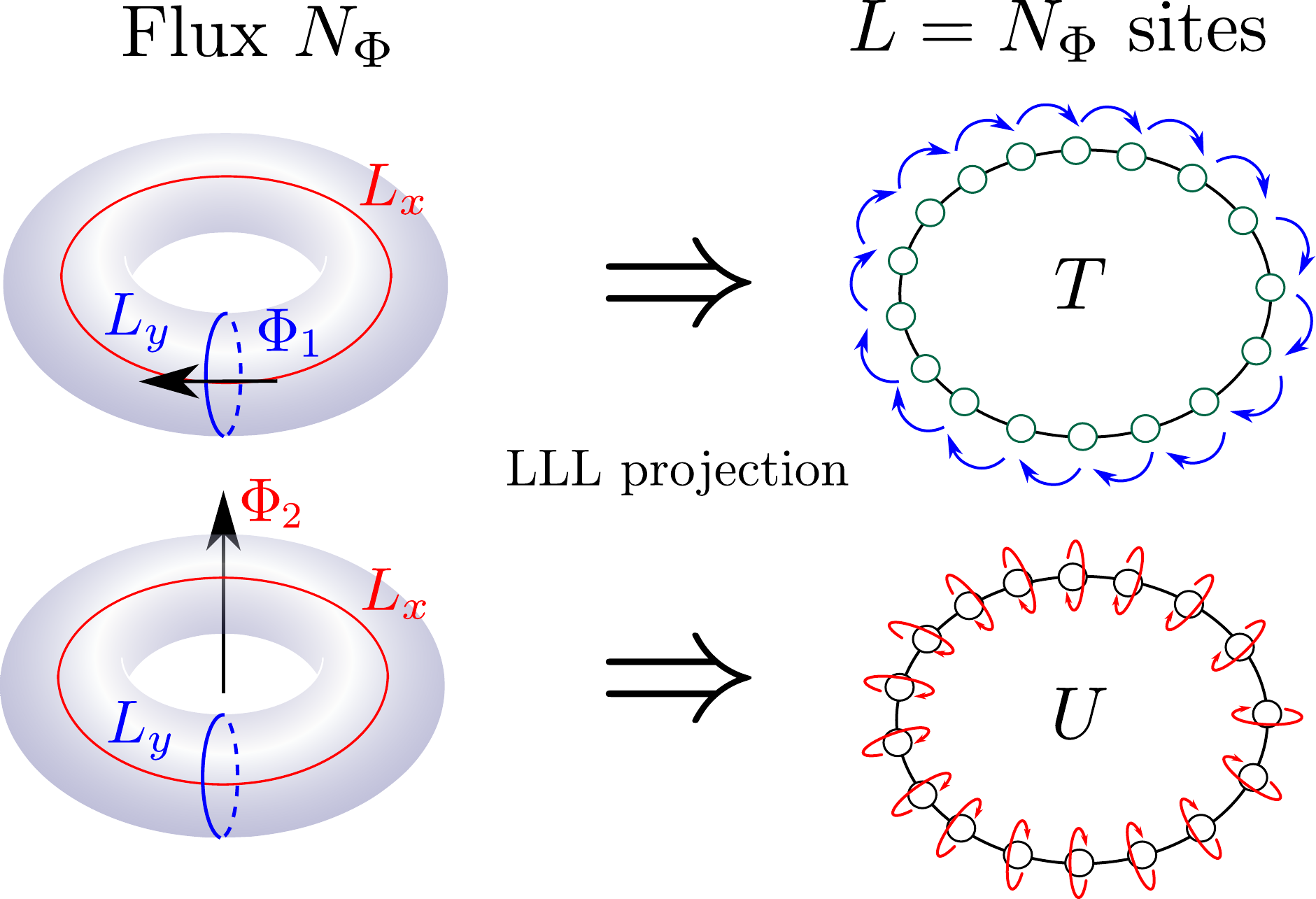} \caption{\textbf{LLL projection and flux insertion in the quantum Hall liquid.}
The LLL projection of a 
quantum Hall liquid on a torus maps the two dimensional state into a 
one-dimensional ring of particles in orbital space. Inserting a flux quanta through one  of the two cycles of the torus (depicted in red and blue) corresponds to a unitary
operation that acts between the different ground states in the torus. In the one dimensional representation, these unitary operations correspond to translation of the 
guiding centers by one orbital ($T$), or multiplication by a phase ($U$), depicted by blue and red arrows, respectively.}
\label{fig:torus} 
\end{figure}

In the quantum Hall context, the operators $D(g)$ of Eq.~(\ref{symmetries}) correspond to inserting fluxes through the two periods of the torus, Fig.~\ref{fig:torus}. In the 1D representation
they are the translation operator $T$ and the operator $U$, that measures the center of mass of the particles in orbital space. They are given by
\begin{equation}
 T=\exp\left\{\frac{2\pi i}{N_\Phi}\sum_{k=0}^{N_\Phi-1}\! k\hat{\tilde{n}}_k\! \right\}; \qquad U=\exp\left\{\frac{2\pi i}{N_\Phi}\sum_{l=1}^{N_\Phi} l \hat{n}_l\!\right\},
\end{equation}
where $\hat{n}_l$ and $\hat{\tilde{n}}_k$ are the number operators at position $l$ and with momentum $k$, correspondingly.
Note that $T$ and $U$ satisfy $T U T^\dagger=e^{-2\pi i \nu}{U}$. 
They thus provide a projective representation of the group $\mathcal{G}=\mathbb{Z}_m\times\mathbb{Z}_m$ with $m=1/\nu$. This group is represented by  $D(g)=U^aT^b$, 
where $a,b=0,\ldots, m-1$, while  $U^m=T^m=\openone$ and $UT=\omega TU$ with $\omega =\exp(2\pi i/m)$.

One can choose a basis $|\Psi_a\rangle$, such that e.g. operator $U$ is diagonal, with its eigenvalues, $\omega^a$, along the diagonal. In this basis, the operator $T$ acts as a raising operator, 
because, if $U|{\Psi}_a\rangle=e^{i\frac{2\pi a}{m}}|{\Psi}_a\rangle$, then $UT|{\Psi}_a\rangle=e^{i\frac{2\pi(a+1)}{m}}T|{\Psi}_a\rangle$.
In the context of the quantum Hall effect,  states $|\Psi_a\rangle$ are the $m$-fold degenerate Laughlin ground states on the torus\cite{Tao1983,Haldane1985}.  

Note that, although the construction of the Lindbladian using the projective symmetry ensures that the eigenvalue $\lambda=0$ of $\mathcal {L}$
is $m$-fold degenerate, the frustration-free condition Eq (\ref{frust_free}) enlarges the degeneracy of $\lambda=0$ to $m^2$. This is seen as follows: The 
symmetry operators $U$ and $T$ ensure that the density matrices $ |\Psi_a\rangle\langle\Psi_a| $ for $ a =0\dots m-1 $ share the same eigenvalue (they are related by conjugation with $T$). In general the matrices 
$|\Psi_a\rangle\langle\Psi_{a+p}|$ are related, for a fixed $p$, by conjugation with $T$. Now these $m$ different families (each labeled by $p =0\dots m-1$) share the same eigenvalue between them due to the 
frustration-free condition which ensures that if $|\Psi_a\rangle$ is annihilated by the quantum jump operators, then all the matrices $|\Psi_a\rangle\langle\Psi_b|$ are as 
well.

{\bf Lindblad operators from Quantum Hall physics.-}
In the narrow torus limit, $\kappa\gg1$, one can truncate expressions for operators $\ell_{s,n}$ in Eq.~(\ref{Q_operator_laughlin}), 
which become short-range in $n$. In this limit we have\footnote{Hereafter we switch from the orbital guiding center index $n$ to the real space site index $i$.}
\begin{equation}
\ell_{0,i}=c_ic_{i+2}\quad \mbox{and} \quad \ell_{1,i}=c_ic_{i+1}+\beta c_{i-1}c_{i+2},
\end{equation}
where $\beta=\eta(\frac{3}{2})/\eta(\frac{1}{2})= 3e^{-2\kappa^2}$. These operators transform as $U \ell_{s,j}U^\dagger = e^{\frac{4\pi i}{N_\Phi}(j+1-\frac{s}{2})}\ell_{s,j} $ and $T \ell_{s,j}T^\dagger= \ell_{s,j+1} $.
Hereafter we regard $\beta$ as an arbitrary  parameter. 
We now employ these results to construct quantum jump operators that drive the system into the frustration free DDS. 
In contrast with the $\ell_{s,i}$ operators in the previous section, here these operators represent processes in a real lattice of $N_\Phi$ sites, 
where $c_i$ destroys a fermion at the lattice site $i$. We assume $m=3$ and the fermion density  $\nu=1/3$. 

We also assume a purely dissipative evolution $\dot{\rho}=\mathcal{L}{(\rho)}$ with $H=0$, Eq.~(\ref{proj_rep_eq}), and the quantum jump operators 
$\tilde{\ell}_{i}= R^\dagger_{i}\mathcal{Q}_{i}$ with $R^\dagger_{i}= c_i^\dagger c_{i+1}^\dagger + t(\ell^\dagger_{0,i}+\ell^\dagger_{0,i+1}) $, and
\begin{equation} \label{QJops}
\mathcal{Q}_{i}= \ell_{1,i}+A(\ell_{0,i-1}+\ell_{0,i+1})+B(\ell_{1,i+1}+\ell_{1,i-1}).
\end{equation}
Here operators $\mathcal{Q}_{i}$ are linear combinations of the operators that annihilate the Laughlin states. 
Parameters $t,B,A$ and $\beta$ are determined by a realization of the dissipative dynamics and are non-universal. The DDS only depends on $\beta$. 
For the purposes of this work we take them as free parameters.

\begin{figure}[ht!]
\includegraphics[width=\linewidth]{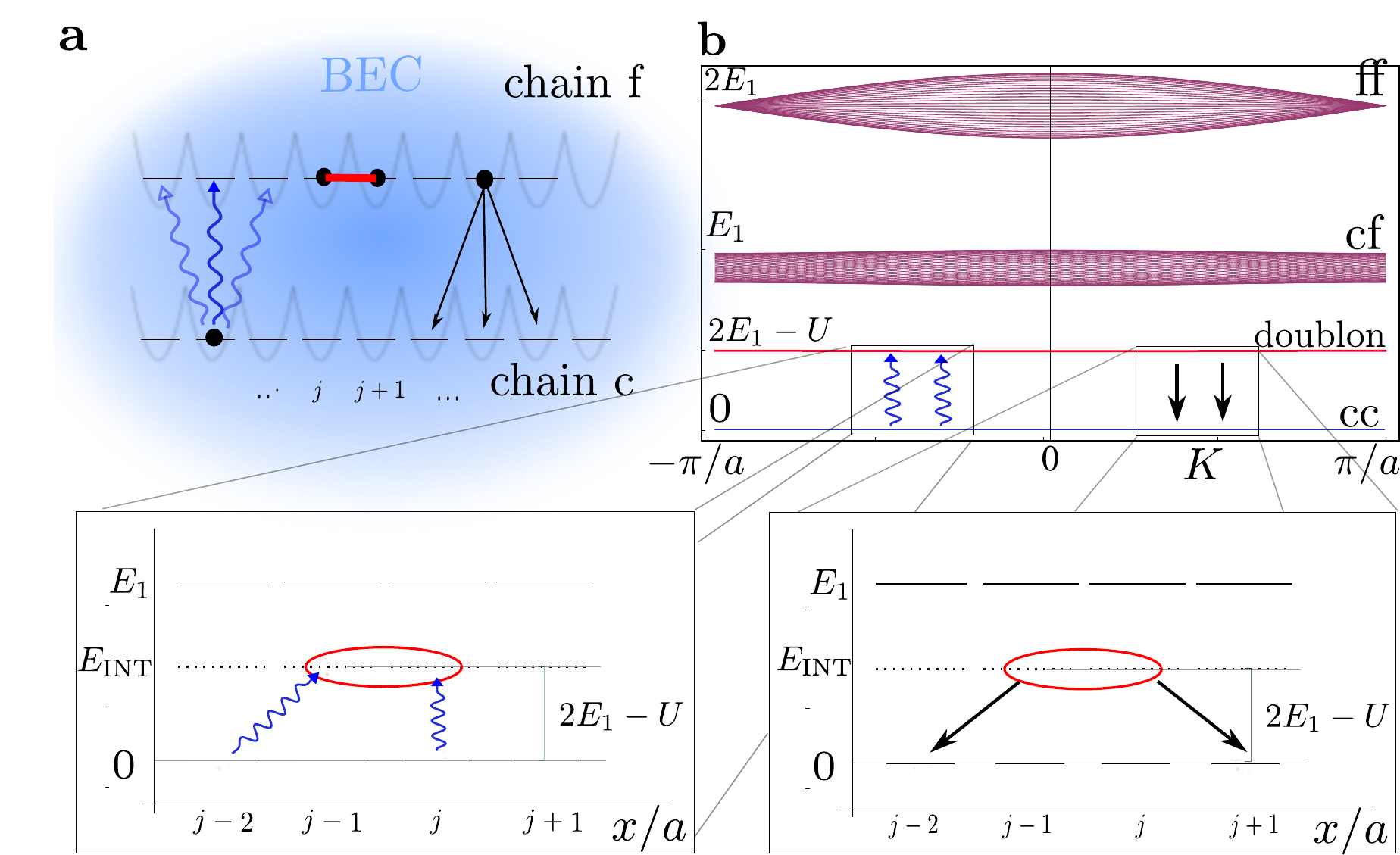} \caption{\textbf{Implementation of Lindblad operators.} a.- Two 
chains c and f, filled with fermions are immersed in a bath, realized as a Bose-Einstein condensate (BEC). Transitions between the two chains, 
that have different
chemical potential are mediated by the absorption of a photon from an external laser (blue wavy lines), or due to the relaxation induced by the bath (black arrows). 
-Particles in the upper band are subject to nearest neighbor attractive interactions, of magnitude $U$ (red line).
b.- Two-particle excitation spectrum, consisting of three free-particle bands (cc, cf and ff) and the {\em doublon} band as function of the total momentum of the pair.  
The laser frequency is red detuned from the transition energy 
$2E_1 -U$, so that the laser mainly creates doublon excitations.  These excitations can decay to the lower band 
by emitting phonons in the bath (lower right panel).}
\label{fig:construction} 
\end{figure}

{\bf Realization of Lindblad operators $\tilde{\ell}_i$.-} The dissipative evolution dictated by the $\mathcal{Q}_i$ operators can be obtained by
coupling two systems: a target fermion chain (where $c_{i}$ destroys a fermion at site $i$) with no intrinsic dynamics ($H_0=0$) and a
fermionic chain with Hubbard interactions described by the Hamiltonian $H_1= \sum_{i}J_1(f_{i}^\dagger f_{i+1}+\text{h.c}) -U\sum_i n_{i}n_{i+1} + E_1 n_{i},$
where  $f_i$ destroys a fermion in this at position i in this chain and $n_i=f^\dagger_i f_i$. We assume that the chemical potential $E_1$ is the largest energy scale in the system. 
These two chains interact through an external classical radiation, with coupling Hamiltonian 
$H_{\rm rad} = \Omega \cos(\omega t)\sum_i f^\dagger_{i}(c_{i}+\alpha(c_{i-1}+c_{i+1}))+\text{h.c}$,
where $\Omega$ is the intensity of the radiation, $\alpha$ parameterizes the spatial laser envelope, and $\omega$ is the frequency of the monochromatic light.
The role of the driving is to excite particles from the target chain to the interacting chain. The particles then relax to the 
lower energy state interacting with the bath, that provides dissipation. These components are shown in Fig. \ref{fig:construction}a

As a motivation for this construction, let's consider how a charge-density-wave (CDW) state decouples from the dynamics in this context, in the limit $J_1<<U$ and $\alpha=0$, 
becoming a dark state. At first order in $\Omega$, exciting a single particle from 
the $c$ to the $f$ chain is strongly suppressed as $E_1$ is large. But in second order ($\frac{\Omega^2}{2E_1-U-\omega}$), the radiation Hamiltonian can create two 
states in the $f$ chain, which can bound in a doublon, consisting on a tightly bound pair of fermions (due to the Hubbard attraction). The wavefunction of the doublon decays 
exponentially with the distance $d$ between its two constituents as $t^d$ with $t\sim J_1/U\ll 1$. This means that, for the laser to create a doublon, the particles in the
$c$ chain should be near each other, as the laser acts locally. In particular, the state which locally contains nearby fermions $c^\dagger_i c_{i+1}^\dagger|0\rangle$ will be
affected by radiation, as well as $c_{i}^\dagger c_{i+2}^\dagger|0\rangle$, where 
$|0\rangle$ is the state with no particles. The first local configuration that is not affected by radiation, becoming {\it dark} is 
$c_{i}^\dagger c_{i+3}^\dagger|0\rangle$, as the fermions are too far away to be excited into a configuration with a non-vanishing matrix element with the doublon. The whole
system will decouple when reaches one of the CDW states. Increasing the range of the laser (by letting $\alpha \neq 0$), creates superpositions.

The Lindblad operators (\ref{QJops}) are obtained considering transitions between the doublon band and the low energy band in the system (more details in the supplemental information).
After performing the rotating wave approximation to account for the time dependence of the radiation Hamiltonian, the dynamics of the system occurs between the lower $c$, 
and the doublon bands (see Fig \ref{fig:construction}b upper panel). Using second order perturbation theory in $\Omega$, we obtain the matrix elements of the transitions 
between the doublon states $|d_i\rangle=f^\dagger_{i}f^\dagger_{i+1}|0\rangle$ and the lower band states $|i,j\rangle=c_{i}^\dagger c_{j}^\dagger|0\rangle$, 
The transition process from lower band to doublon reads (after adiabatic elimination of the doublon)
$\tilde{\ell}_i = A_\Omega^2 c_i^\dagger c_{i+1}^\dagger\mathcal{Q}_{i}$, valid for $t=\alpha^3\ll1$, with $\mathcal{Q}_i$ given in (\ref{QJops}).
The prefactor $A_\Omega\sim \Omega^2/(2E_1-U-\omega)$, while $A,B,$ and $\beta$ entering the definition of $\mathcal{Q}$ satisfy $A =\alpha, B=\alpha A_\Omega$ and $\beta = \alpha^2$. 
The fermion operators in $\mathcal{Q}_i$ all act on the chain $c$. Finally, taking into account the transitions from the doublon back to the $c$ chain,
which is mainly mediated by the dissipation with the bath, and integrating out the doublon states, we arrive at the Lindblad operators (\ref{QJops}).

{\bf Structure of the DDS.-}
Heuristically, one can understand the roles of $R^\dagger_i$ and $\mathcal{Q}_{i}$ as follows. Operator $\mathcal{Q}_{i}$ checks if at the site $i$ the state  matches the 
local configuration of one of the Laughlin-like states. If true, it gives zero and the system stops evolving locally; if false, the operator 
$R^\dagger_i\mathcal{Q}_{i}$ scrambles the particles. As long as this process can efficiently mix the particles locally, all states in the Hilbert space 
eventually evolve into  DDS, spanned by the three Laughlin states. Crucially, the decay into these states is a consequence of the projective symmetry, enforcing existence of the degenerate space with $\mathcal{Q}_{i}{\bm\rho}=0$ and thus $\dot{\bm\rho}={\cal L}({\bm\rho})=0$. 

The basis of such DDS is formed by the Laughlin states $|\Psi_a\rangle$, where $a= 0,1,2$, which  are annihilated by all composite operators, $\ell_{s,i}|\Psi_a\rangle=0$,   
for all $s,i$ 
(and thus by the quantum jump operators). Assuming periodic boundary conditions, 
these states are given by\cite{Nakamura2012,Ortiz2013} MPS\cite{Fannes1992,Klumper1993}
\begin{equation}\label{MPS_state}
|\Psi_a\rangle=\mathcal{N}\,{\rm tr}\big\{g^{a}_1g^{a}_1\dots g^{a}_{L/3}\big\},
\end{equation}
 where $\mathcal{N}$ is a normalization factor, $a= 0,1,2$ and
\small
\begin{eqnarray}\nonumber
g^{0}_i  &\!=\!&  
\left(\!\! \begin{array}{cc}
|{ \circ\circ\bullet}\rangle_i & |{ \circ\circ\circ}\rangle_i\\
-\beta|{ \bullet\bullet\circ}\rangle_i & 0
\end{array}\!\!\right)\!;\qquad
g^{1}_i  \!=\!  
\left(\!\! \begin{array}{cc}
|{ \bullet\circ\circ}\rangle_i & |{ \circ\bullet\bullet}\rangle_i\\
-\beta|{ \circ\circ\circ}\rangle_i & 0
\end{array}\!\! \right)\!;\\
 g^{2}_i & \!=\!&  
\left(\!\!\begin{array}{cc}
|{ \circ\bullet\circ}\rangle_i & |{ \circ\circ\bullet}\rangle_i\\
-\beta|{ \bullet\circ\circ}\rangle_i & 0
\end{array}\!\!\right).
\end{eqnarray}
\normalsize
The state $|{ \circ\circ\circ}\rangle_i$ represents the three consecutive empty sites at positions $(3i-2,3i-1,3i)$, while 
a full dot represents an occupied site, e.g. $|{ \bullet\circ\circ}\rangle_i=c^\dagger_{3i-2}|{ \circ\circ\circ}\rangle_i$, {\em etc}. 
The dark space basis vectors, $|\Psi_a\rangle$, are related by a translation $T$ by one site, as $T|\Psi_a\rangle=|\Psi_{a\oplus 1}\rangle$, where $\oplus$ is an addition modulo 3. In the basis $|\Psi_a\rangle$, the operators $T$ and $U$ are represented by the $3\times 3$ matrices 
\begin{equation}
 {T}= \left(\begin{array}{ccc}
0 & 1 & 0 \\
0 & 0 & 1\\
1 & 0 & 0
\end{array}\right);\quad\quad  {U}= \left(\begin{array}{ccc}
1 & 0         & 0\\
0 & \omega^2  & 0 \\
0 & 0         & \omega
\end{array}\right); \quad \omega=e^{\frac{2\pi i}{3}}.
\end{equation}
Within DDS the density matrix  is  $\bm{\rho}=\sum\limits_{a,b=0}^{2}\varrho_{ab}|\Psi_a\rangle\langle \Psi_b|$, with
$\varrho_{ab}$ a $3\times 3$ positive semidefinite, hermitian matrix with unit trace. In general, the structure of the density matrix within DDS depends on 
initial conditions, which determine the parameters $\varrho_{ab}$.

The basis vectors that define the DDS depend explicitly on the parameter $\beta$. For $\beta=0$ the DDS is spanned by a mixture of three different classical CDW \cite{Tao1983}, e.g. 
$|\circ\bullet\circ\circ\bullet\circ\dots\rangle$, with periodic density and sharp structure factor in each basis vector.  Changing this parameter modifies the average local density. 
The latter is given by
\begin{equation}
 {\rm Tr}({\bm \rho}\,\hat{n}_{3i+a})=\frac{3\varrho_{aa}-1}{2\sqrt{1+4|\beta|^2}}+\frac{1}{2}(1-\varrho_{aa}),
\end{equation}
indicating that, if $\beta$ is known, the local density measurements are enough to determine the probabilities $\rho_{aa}$.

\begin{figure}[ht!]
\includegraphics[width=\linewidth]{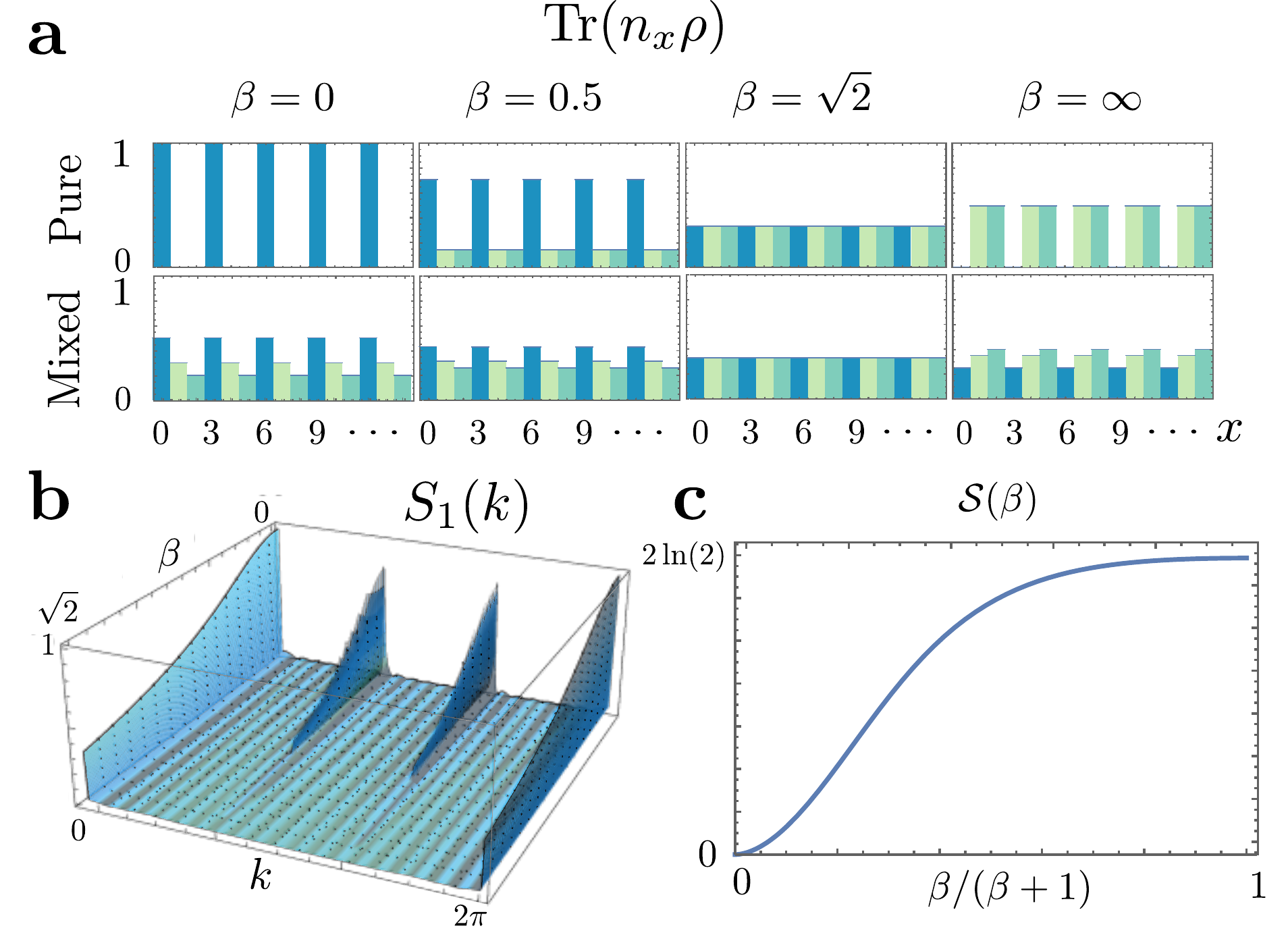} \caption{\textbf{Characteristics of the DDS basis states}
a.- Expectation value ${\rm Tr}({\bm \rho}\,\hat{n}_{x})$ for different positions $x$. 
Starting from a CDW configuration,  the system is evolved using a protocol with a given $\beta$. 
This generates a DDS state that depends on $\beta$. The top row represents the pure state $\varrho_{00}=1$, while the second row represents the mixed state
$\varrho_{00}=0.5,\varrho_{11}=0.3,\varrho_{22}=0.2$. Different colors are used to help track the changes in average occupation at each site and to highlight the 3-site 
periodicity of the density. b.- Static structure factor $S_1(k)$ for different values of $\beta$, for a pure state with $\rho_{11}=1$. At $\beta=0$ the system
is in the CDW state, with a definite spatial periodicity, indicated by the peaks in 
$S_1(k)$ at $k=0,\frac{2\pi}{3},\frac{4\pi}{3}$. Increasing $|\beta|$, the system becomes more homogeneous.
c.- Entanglement entropy of the DDS as a function of $\beta$.}
\label{fig:density} 
\end{figure}

An alternative way of characterizing the DDS is through the correlation function of local observables. To highlight the relation with the CDW states, we focus on the static structure factor 
\begin{equation}\label{eq:densitydensityFT}
 S_a(k) = \frac{3}{L}\sum_{i=1}^L {\rm Tr }( \bm \rho\, \hat n_{a} \hat n_{a+i-1}) e^{i k(i-1)},
\end{equation}
shown in Fig. \ref{fig:density}b. As $\beta$ increases, the system transitions from a crystal-like state with the well defined translational symmetry of three sites, into a 
more homogeneous state, where the density is uniform across the system.

Finally, to describe the quantum nature of the DDS basis states, we compute their entanglement entropy. We separate the degrees of freedom of the system in two 
complementary large regions $A$ and $A^c$ and define the partial density matrix $\rho_A =\rm Tr_{A^c}(|\Psi_a\rangle\langle \Psi_a|)$. 
The entanglement entropy is then 
$\mathcal{S}(\beta)=-{\rm Tr}({\rho}_A\ln({\rho}_A))$. The result is shown in Fig. \ref{fig:density}c. We observe that the entanglement entropy is monotonic with $\beta$. It reaches its  maximum value of $2\ln(2)$ for MPS of bond dimension 2 at $|\beta|\rightarrow \infty$.

{\bf Time evolution and global diagnostics.-}
Now that we have constructed a dissipative evolution that drives the system into the DDS, we discuss how the system approaches the DDS. 
We analyze the Lindbladian evolution with the quantum jump operators (\ref{QJops}) numerically. 
The decay into the DDS is evaluated using a quench protocol: the system is initiated in the CDW state $|\circ\bullet\circ\circ\bullet\circ\dots\rangle$, 
which is one of the dark states of the Lindbladian at $\beta=0$. This state is evolved then using the Lindblad operators (\ref{QJops}) with $A=B=t=1$ and $\beta \in[0,1]$ for simplicity 
(the results are qualitatively similar for slightly varying these parameters). 
 In order to obtain the evolution of the system we perform exact diagonalization using Runge-Kutta (RK) integration\cite{Numerical_Recipes} of the master equation, for 
 systems of sizes up to $L = 15$.  

\begin{figure}
 \centering
  \includegraphics[width=0.45\textwidth]{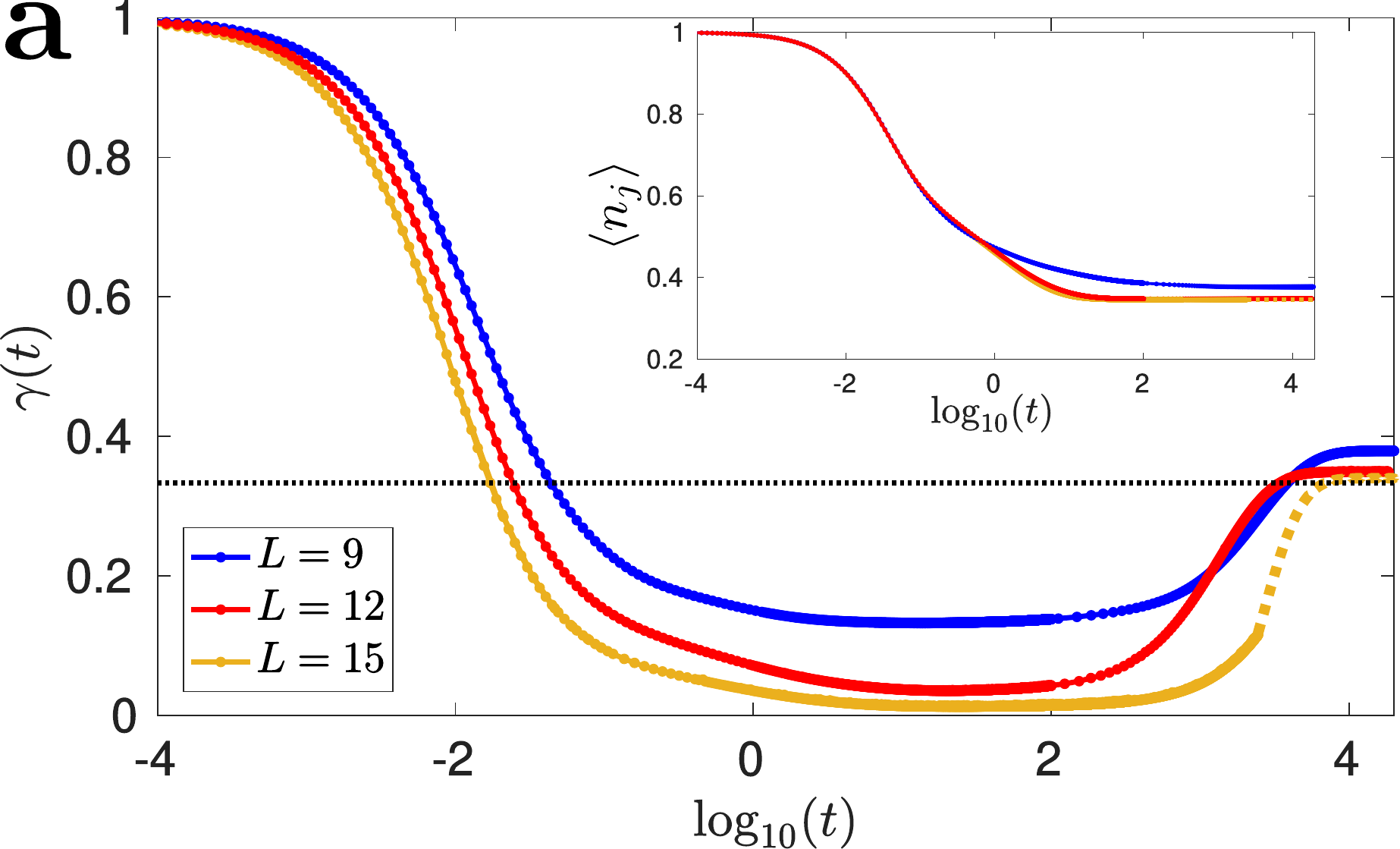}
  \includegraphics[width=0.23\textwidth]{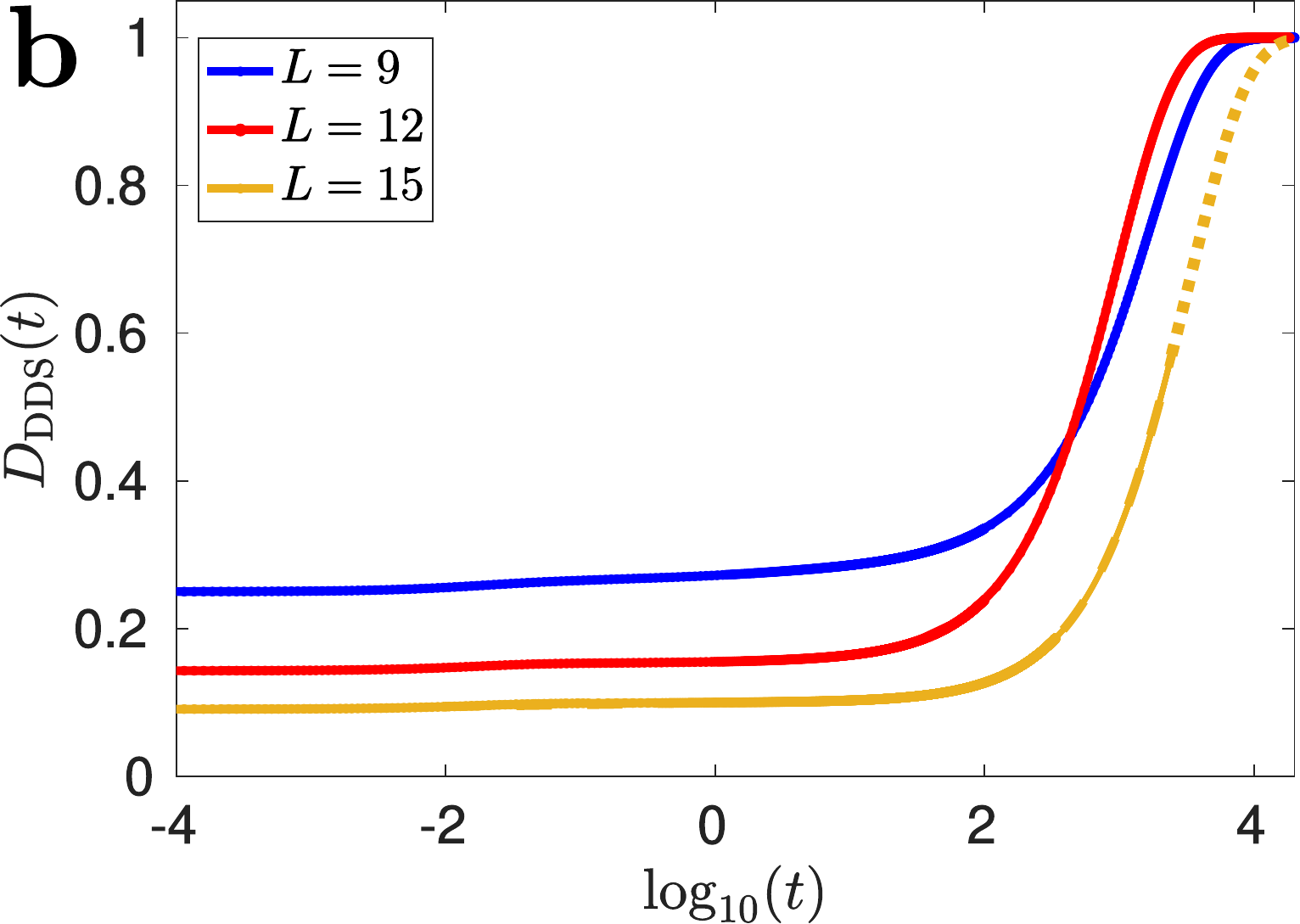}
  \includegraphics[width=0.23\textwidth]{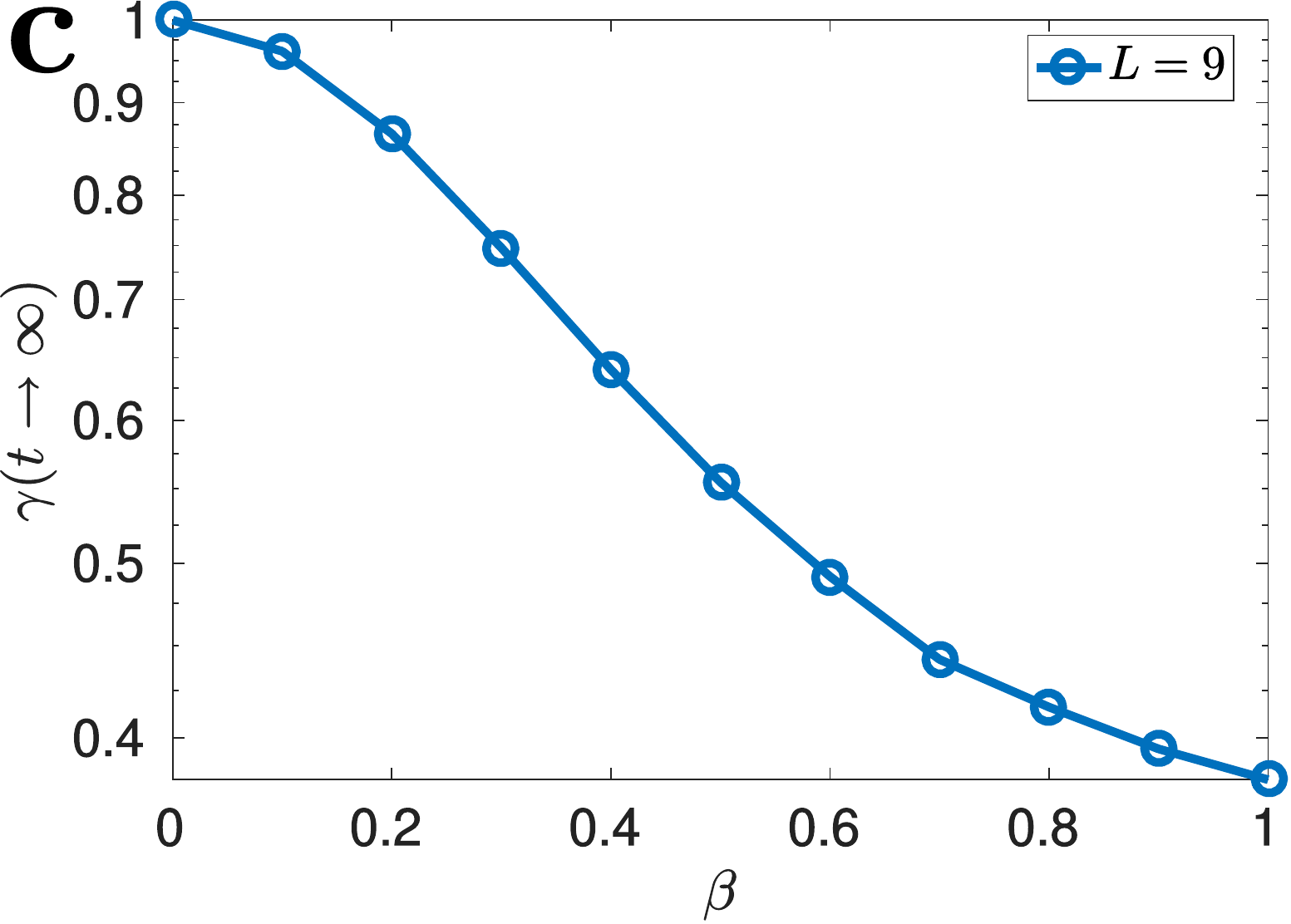}
 \caption{\textbf{Time evolution and decay into the DDS.}
 a.- Evolution of the purity $\gamma(t)$, for different sizes. Starting from a pure CDW state, the system becomes highly mixed
 before starting leaking into the DDS. The dotted line shows the minimum purity possible in the DDS, corresponding to a fully mixed state.
 Inset.- Evolution of a local observable (density). We observe that the local density relaxes to it value in the DDS in a shorter timescale compared with the 
 time required for the system to enter the DDS. This happens independently of the system size.
 b.- Approach of the DDS, spanned by the Laughlin-like states on the narrow torus, for different system lengths $L$.
 c.- Purity in the DDS as a function of $\beta$. The purity of the DDS depends strongly
 on $\beta$, indicating that the purity can be maintained if the final state is close to the initial one.
 The dotted lines for the $L=15$ case for the purity and density evolution are obtained from an approximation for the Lindbladian on a smaller subspace \cite{Note2}. 
 }\label{fig:purity}
\end{figure}

To characterize the steady state mixture, we compute the purity of the state, defined as $\gamma(t) = {\rm Tr}\{\rho^2(t)\}$. From Fig. \ref{fig:purity}a, we find that the purity approaches $1/3$ with larger system sizes.  
This is indeed the case for a sudden quench from $\beta(t\leq 0)=0$ to $\beta(t>0)=1$. In this scenario, the system explores an extensive portion of the Hilbert
space, becoming highly mixed, as seen in the intermediate region of Fig. \ref{fig:purity}a, where the purity plateaus at a minimum. Only after the system is sufficiently 
mixed, it starts  approaching DDS and its purity increases. The information about the initial state is practically lost in the intermediate mixing process, and the eventual steady state is  a highly mixed state within DDS. 

Convergence to DDS, spanned by Laughlin-like dark states 
$\{|\Psi_a\rangle\}_{a=0,1,2}$,   may be visualized  by   
 \begin{equation}\label{eq:dness}
  D_{\rm DDS}(t) = {\rm Tr }\{ \rho(t) \mathcal{P}_{\rm{DDS}} \},
 \end{equation}
where  $\mathcal{P}_{\rm{DDS}} = \sum_{a=0}^2 |\Psi_a\rangle \langle \Psi_a|$ is a  projector onto the DDS.  Figure \ref{fig:purity}b shows that the system indeed evolves towards
the Laughlin-like DDS, proving that this is the only non-decaying subspace of the Lindbladian evolution.  
At large times,   one finds $1- D_{\rm DDS}(t)\propto e^{-\lambda_{0}t}$, where the rate $\lambda_{0}$ is given by the lowest non-zero eigenvalue of the Lindblad operator. 

One notices that $\lambda_0$ is slowly decreasing upon increasing  the system size.  
We note that local observables, like the particle density, fast approach their steady state values in a way  which is 
independent of the system size, Fig.\ref{fig:purity}a. This separation of scales indicates that, while locally the system reaches a configuration that is close to the dark states that span the DDS, globally it takes much longer to
fully reach the DDS.

If instead of quenching the system into $\beta=1$, we quench it into $\beta\ll 1$, 
we observe a very different behavior.
Here the system does not have to explore an extensive part of the Hilbert space before it reaches the DDS.  
As a result, the purity remains close to 1 at all times, as can be seen in Fig. \ref{fig:purity}c.

\begin{figure}[hbt!]
 \centering
 \includegraphics[width=0.45\textwidth]{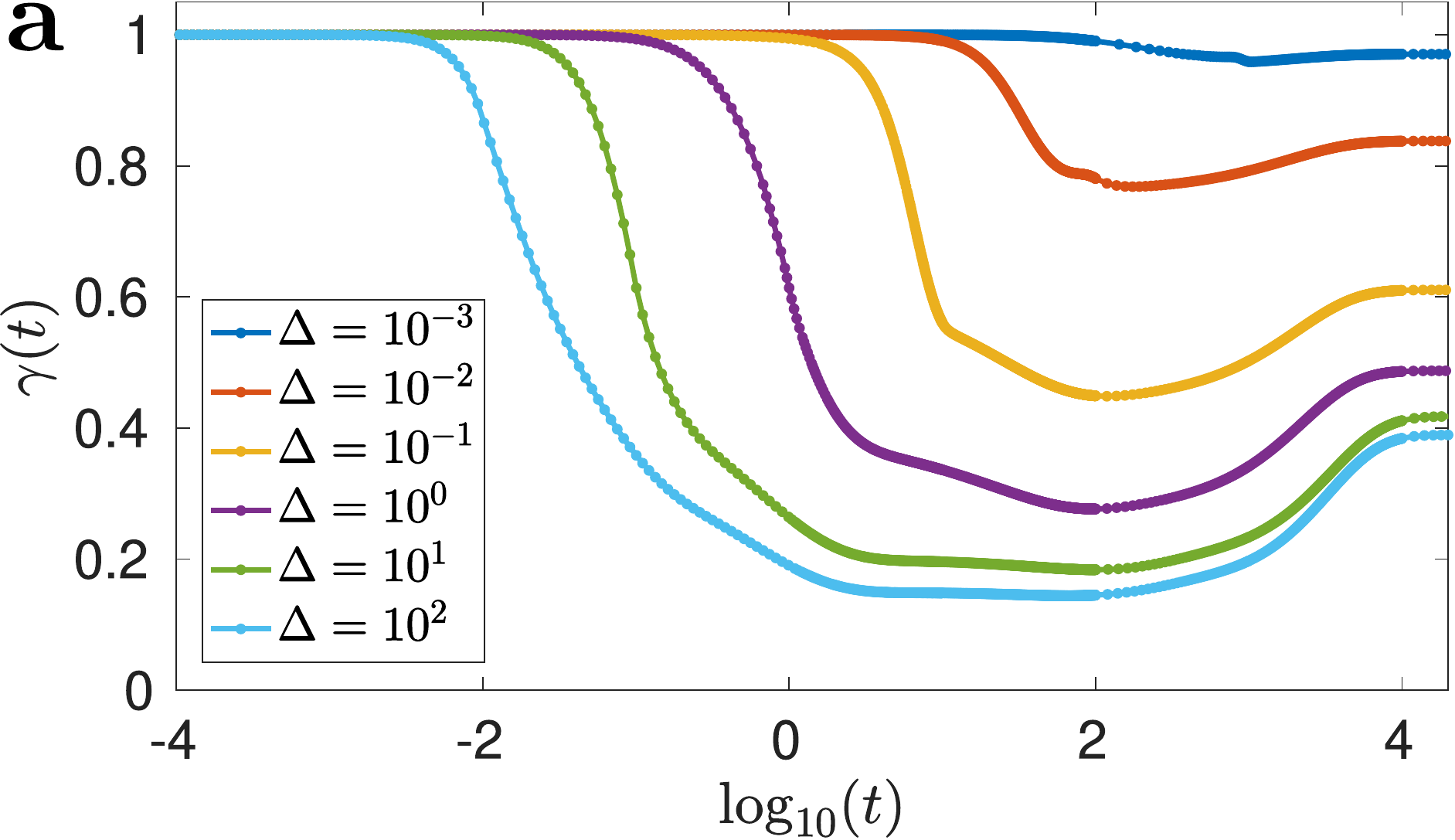} 
 \includegraphics[width=0.235\textwidth]{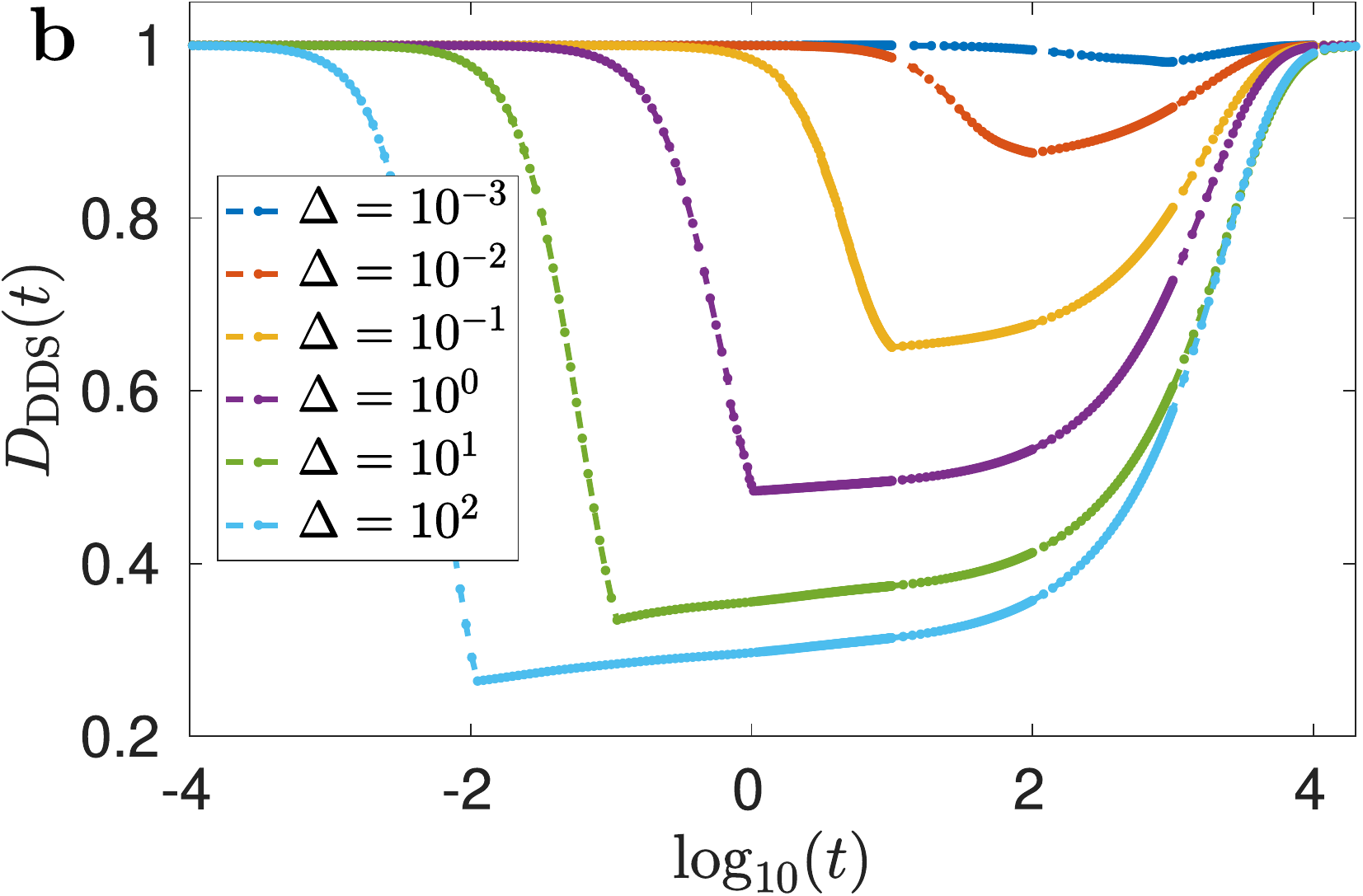}
 \includegraphics[width=0.235\textwidth]{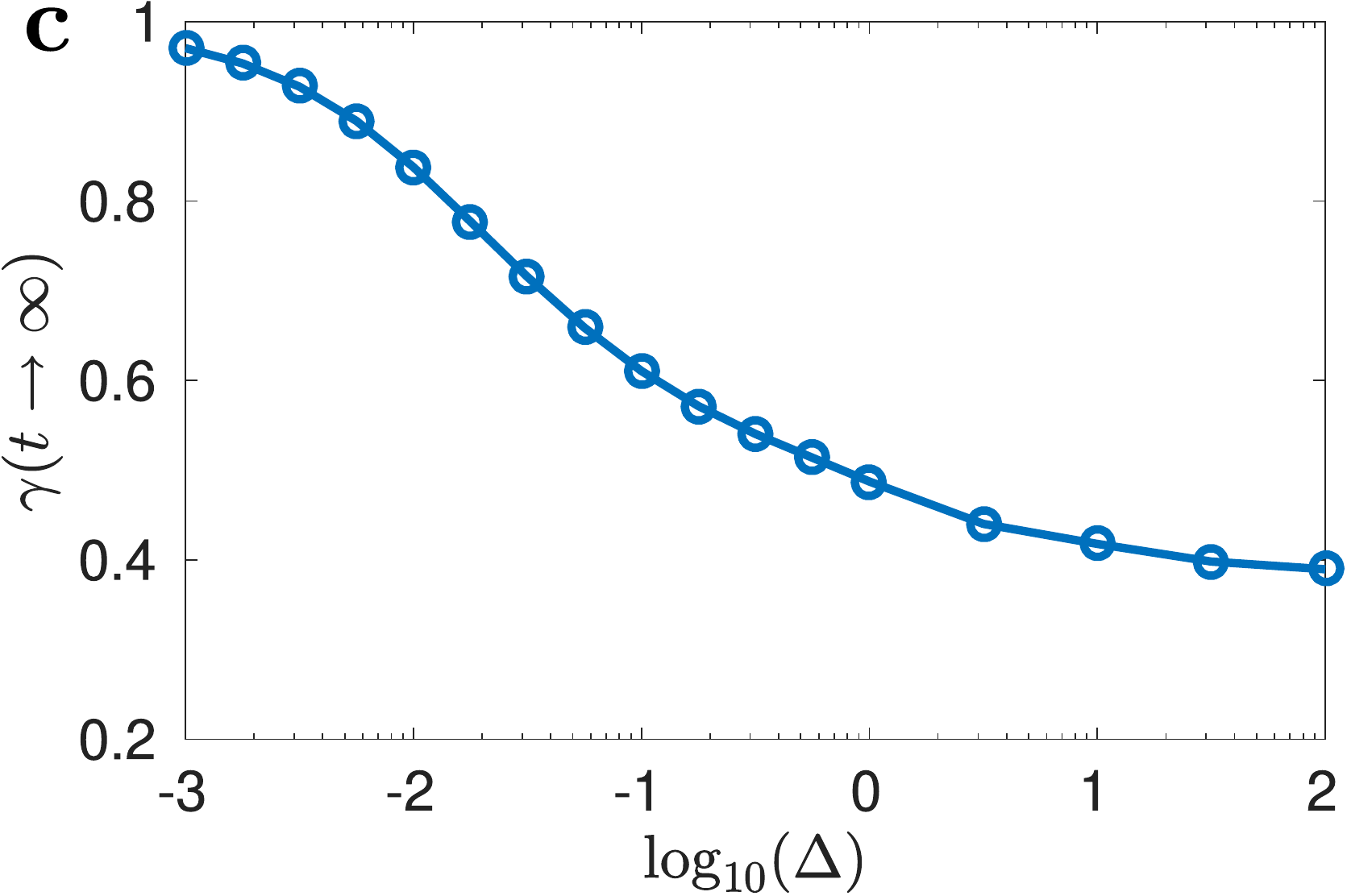} 
 \caption{\textbf{Adiabatic evolution of the system.}
a.- The purity of the final state for different ramp velocities $\Delta$ in a system with $L = 9$ sites.
A slower variation in the adiabatic protocol leads to a higher purity. b.- The system ends up
in the DDS regardless of the ramp velocity. Different $\Delta$'s control how much of the Hilbert space is explored. c.- The purity
of the final state can be manipulated via the ramp velocity.}\label{fig:adiabatic_purity}
\end{figure}

{\bf Adiabatic evolution.-} Although the previous analysis shows that the system does not generically end up in a pure state, it is possible to
increase the purity of the final state by performing an adiabatic evolution from a pure state \cite{Avron2012, Sarandy, Albert2016}.
To illustrate this, we evolve the system from an initial state given by a superposition of the three CDW configurations. This allows us to characterize the coherences in the MPS basis throughout the 
adiabatic evolution (sec. III in SI). Individual CDW states can be created using existing experimental techniques\cite{Schreiber2015}. We then evolve this state with the Lindblad operators (\ref{QJops}) using a time-dependent $\beta$ parameter: $ \beta(t)= \Delta \cdot t$ for  $0<t\leq 1/\Delta$ and $\beta(t)=1$ for $t>1/\Delta$, 
where $\Delta$ is the ramp velocity.  The purity of the final state depends on $\Delta$ as shown in 
Fig \ref{fig:adiabatic_purity}a. For small enough $\Delta$, the system does not explore the whole Hilbert space, but instead remains almost pure throughout its entire evolution.
This mechanism can be used to achieve a purity arbitrarily close to unity. For larger ramp velocities, the system rapidly departs from the initial state, exploring the 
many-body Hilbert space, as shown for intermediate times in Fig.\ref{fig:adiabatic_purity}b, before leaking back to the DDS, which remains the 
only attractor of the dynamics. This increases the departure of the steady state from a pure state and erases the information about the initial state. The steady state
purity as a function of the ramp velocity is shown in Fig.\ref{fig:adiabatic_purity}c.

{\bf Conclusions.-}
We have shown that to achieve a DDS the Lindbladian evolution should have an underlying symmetry, admitting a projective representation. The period of its 2-cocycle determines the dimensionality of the dark space.   
Reaching a DDS, protected against environmental influence, offers a way of maintaining quantum information. 
To manipulate this information, it is necessary to have an access  to high purity states within the DDS.
We found that an adiabatically varying Lindblad operator allows to reach such nearly-pure, entangled states. 
We have demonstrated these ideas by studying the thin torus limit of the $\nu=1/3$ fractional quantum Hall state of matter. Being able to generate and manipulate states 
within a DDS may be utilized for quantum information processing platforms. The many-body nature of the state renders it less fragile against local disturbances.

{\bf Acknowledgments.-} We are indebted to R. Fazio for valuable discussions. 
R.S.  acknowledges  funding  from  by  EPSRC grant EP/M02444X/1, and the ERC Starting Grant No.678795 TopInSy.
F.I. acknowledges the financial support of the Brazilian funding agencies National Council for Scientific and Technological Development - CNPq (Grant No.308205/2019-7) and FAPERJ (Grant No.E-26/211.318/2019).
A.K. was supported by NSF grant DMR-1608238. 
Y.G. was supported  by the Deutsche Forschungsgemeinschaft (DFG) TRR 183 (project B02), and EG 96/13-1, and by the Israel Science Foundation.

\renewcommand{\thetable}{S\arabic{table}}
\renewcommand{\theequation}{S\arabic{equation}}
\renewcommand{\thefigure}{S\arabic{figure}}

\onecolumngrid

\section*{Supplemental Information for:\\ 
Multidimensional dark space and its underlying symmetries: towards dissipation-protected qubits}

In this section we provide supplemental information not been inserted in the main text. We discuss the analysis of the Lindbladian gap of the dissipative model with and without perturbations, as well as an 
expanded analysis of the adiabatic evolution. We revisit the map of the Laughlin state from two dimensions to the orbital basis, where it becomes effectively one dimensional. Finally we include a longer
discussion about the experimental realization of the Lindblad operators.

The dissipative evolution $\dot{\rho} = \mathcal{L}(\rho)$ with $\hat H = 0$ studied in the main text is defined by the quantum jump operators 
$\tilde \ell_{i} = R^\dagger_i \mathcal{Q}_{i}$ for $i=1,...,L$ 
with $R^\dagger_i = \cd{i}\cd{i+1} + t (\ell_{0,i}^\dagger + 
\ell_{0,i+1}^\dagger)$ and 
\begin{eqnarray}
\mathcal{Q}_{i} &=& \ell_{1,i} + A (\ell_{0,i-1} + \ell_{0,i+1})
+ B (\ell_{1,i+1} + \ell_{1,i-1}) \nonumber \\
\ell_{0,i} &=& c_{i} c_{i+2}, \qquad \qquad 
\ell_{1,i} = c_{i} c_{i+1} + 
\beta c_{i-1} c_{i+2}.
\label{eq.lind.operators}
\end{eqnarray}

In Sec.~\eqref{sec.Lindbladian gap} we show our results for the Lindbladian gap in finite system sizes with no perturbations, while the case of the dissipative evolution with imperfections are presented in Sec.~\eqref{sec.Lindbladian.perturbations}. In Sec.\eqref{sec.Adiabatic evolution} we show details of the adiabatic evolution analysis.

\section{Lindbladian gap in finite system sizes}
\label{sec.Lindbladian gap}

In this Section we show our analysis for the dissipative gap of the Lindbladian for finite system sizes. We obtain the gap in two different forms: (i) directly by exact diagonalization of the Lindbladian superoperator, 
or (ii) indirectly by the asymptotic decay rate (ADR) of the quantum state dynamics. While 
exact diagonalization allows us to study the Lindbladian gap for sizes up to $L \sim 12$, the asymptotic decay rate analysis allows the study of larger $L \sim 15$ system sizes.

The exact diagonalization analysis is performed by first describing the Lindbladian superoperator as a linear operator in the extended Hilbert space,
\begin{equation}
 \mathcal{L} \rightarrow |\mathcal{L}\rangle = -i (\mathbb{I} \otimes \hat H - \hat H^T \otimes  \mathbb{I}) + \sum_i \hat \ell_i^* \otimes \hat \ell_i - \frac{1}{2} \mathbb{I} \otimes \hat \ell_i^\dagger \hat \ell_i - 
 \frac{1}{2} (\hat \ell_i^\dagger \hat \ell_i )^T \otimes \mathbb{I}
 \label{eq:vectorized.Lind}
\end{equation}
and then diagonalizing the vectorized Lindbladian $|\mathcal{L}\rangle$. The eigenvalues $\lambda_j$ of the Lindbladian have non-positive real values, with the DDS described by those with zero eigenvalue. The gap of 
the Lindbladian corresponds in this way to the eigenvalue with largest nonzero real part. Ordering the eigenvalues according to their real part $\Re(\lambda_j) \geq \Re(\lambda_{j+1})$, for $j=1$ to the extended 
Hilbert space dimension, the dissipative model of this work have in this way $\lambda_{j} = 0$ for $j=1,...,9$ and the gap is described by the eigenvalue $\lambda_{10}$.

On the other hand, one can also extract the gap by studying the dynamics of the quantum state. In the long time limit only the slower decaying modes of the Lindbladian are relevant to the dynamics and the expectation 
value of any observable $O(t) = Tr(\hat O \hat \rho(t))$ is approximated by
\begin{equation}
 O(t) - O(t \rightarrow \infty) \approx e^{\lambda_{\rm ADR} t}
\end{equation}
where $\lambda_{\rm ADR} < 0$ corresponds to the slower decay mode (or asymptotic decay mode) for the observable $\hat O$.

\begin{figure}
\includegraphics[scale=0.5]{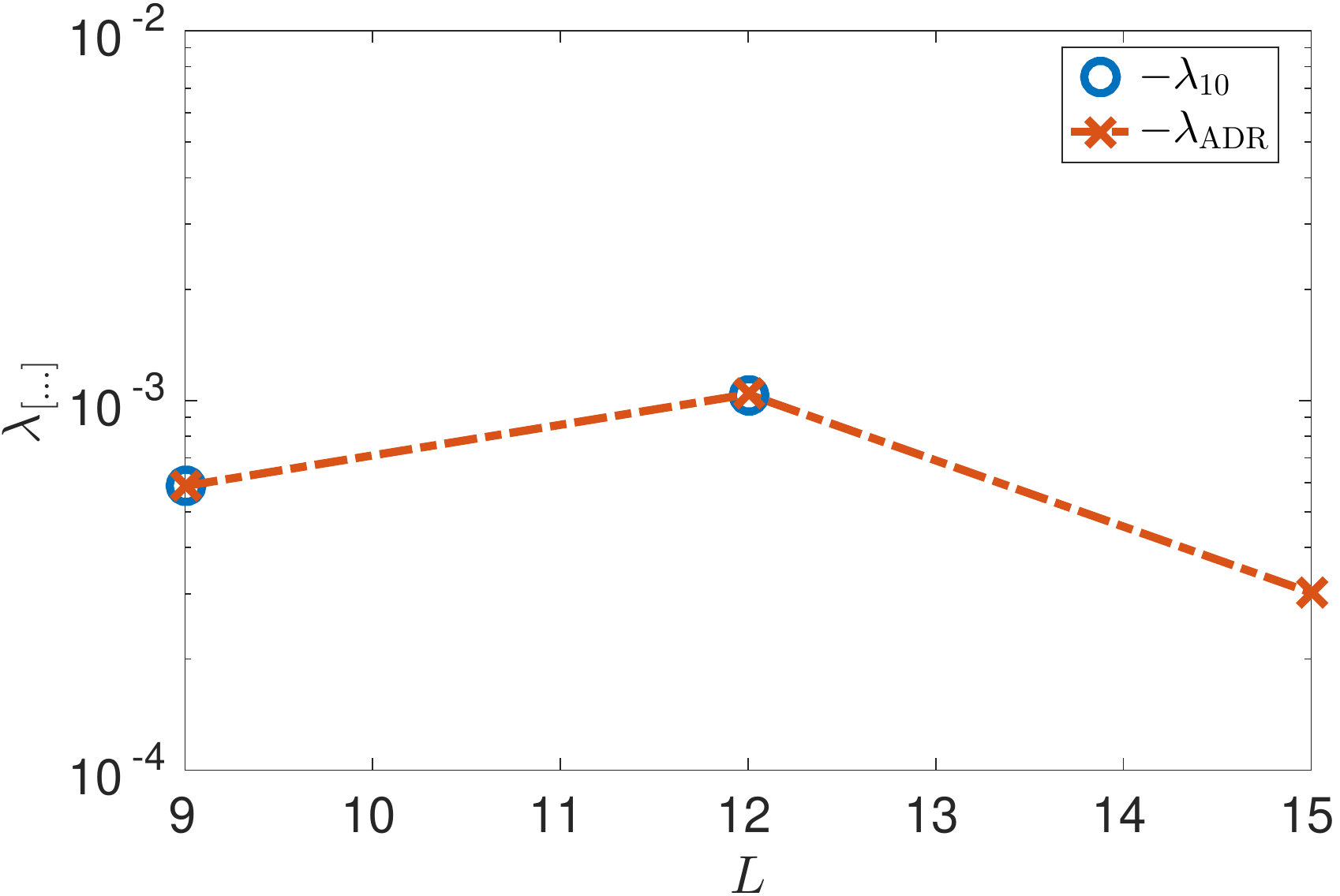}
\includegraphics[scale=0.5]{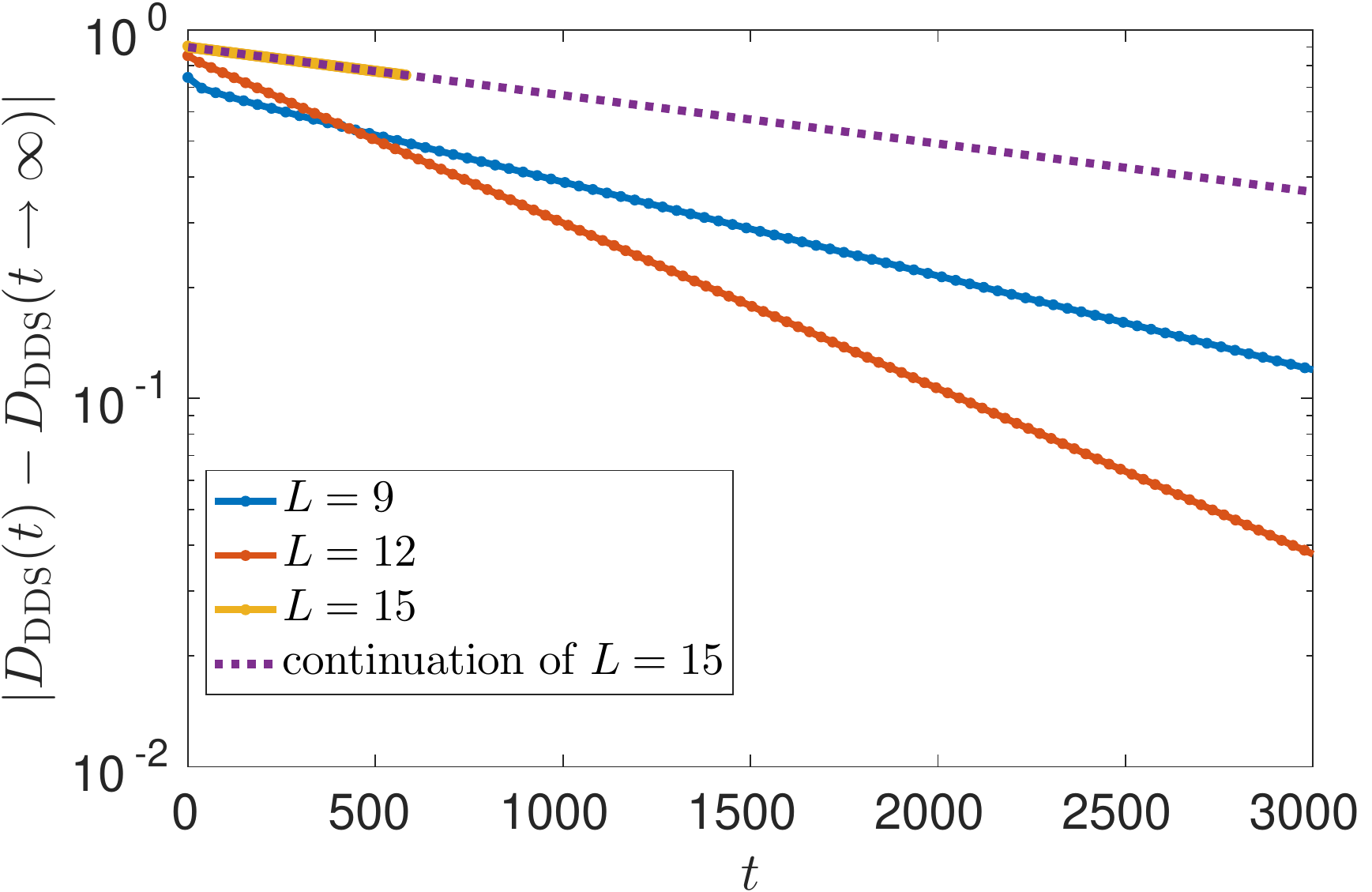}
 \caption{\textbf{Lindbladian gap for finite system sizes.} In the \textbf{left-panel} we show our results for the Lindbladian gap ($\text{gap} = \lambda_{10}$) obtained from exact diagonalization of the vectorized 
 Lindbladian (Eq.\eqref{eq:vectorized.Lind}), and the asymptotic decay rate extracted from the dynamics of $D_{\rm{DDS}}(t)$, for varying system sizes. In the \textbf{right-panel} we show the dynamics of 
 $D_{\rm{DDS}}(t)$ for a quench dynamics starting from an initial CDW state. We show the dynamics in a log-linear scale highlighting the exponential asymptotic behavior towards the steady state value 
 $D_{\rm{DDS}}(t \rightarrow \infty) = 1$. The dashed line corresponds to the expected long time dynamics for the $L=15$ case, according to its ADR coefficient. The Lindbladian parameters in all figures are 
 $A=B=t=\beta=1$.
 }
 \label{fig:lindbladian.gap}
\end{figure}

In Fig.\eqref{fig:lindbladian.gap} we show our results for the Lindbladian gap and ADR analysis of the ${\rm DDS}(t)$ dynamics. 
We observe that the gap from exact diagonalization matches the one obtained from ADR. Furthermore we see that the dissipative
gap increases going from $9$ to $12$ sites, and decreases again for $15$ sites (to slightly the same value as $9$ sites), which is the limit size that we can achieve. Although instructive, these numerical results for 
small systems do not allow us to unequivocally determine the nature of the dissipative gap in the thermodynamic limit. It is suggestive, however, for the presence of either (i) a gapped Lindbladian in the 
thermodynamics limit, or (ii) a gapless Lindbladian with a slow decaying of the gap with system size, excluding e.g. the possibility of an exponential closure of the gap with system size which would be detrimental for 
the preparation and manipulation of DDS in quantum information tasks.

In the main text, we showed that the existence of the symmetry operators $U$ and $T$ ensures that all of the eigenvalues of the Lindbladian are at least m-fold degenerate. It is good to stress that the eigenmatrices of the 
Lindbladian do not translate directly into density matrices. In particular, this means that the symmetry operators $U$ and $T$ do not generate directly a symmetry on the density matrices. The Lindbladian is a linear 
operator acting on the extended Hilbert space $ V=\mathcal{H}\otimes\mathcal{H} $, Eq. \ref{eq:vectorized.Lind} which for concreteness we can take to be spanned by the vector states $|i\rangle|j\rangle$, each defined within a copy of 
$ \mathcal{H}$ . The vectorized density matrices form a subspace $S\subset V$ which is not a vector space, as the sum of two elements of $S$ does not generically belong to $S$ (e.g. given $\rho_{1,2}$ two positive definite 
operators with unit trace, the linear combination $\alpha \rho_1 + \beta \rho_2$  with $\alpha,\beta$ complex numbers is not necessarily positive definite).

\section{Lindbladian perturbations}
\label{sec.Lindbladian.perturbations}
In this Section we study the effects of Lindbladian imperfections on the DDS.
We consider the following different imperfections:

 $\bullet$ Additional set of dephasing  dissipative channels:
 \begin{eqnarray}
 \hat \ell_{j,\rm dephasing} = \sqrt{\epsilon_{\rm deph}} \left(2 \hat c^{\dagger}_j \hat c_{j} - 1\right), \qquad j=1,...,L \label{eq:lind.imp.dephasing}
 \end{eqnarray}
 describing fluctuations of on-site energies, which tend to suppress coherences between classical particle number basis of the fermionic system.
 
$\bullet$ Imperfections in the current Lindblad operators of the form of extra hopping terms:
\begin{eqnarray}
\tilde \ell_j \rightarrow \hat \ell_{j,\epsilon} = \tilde \ell_j + \sqrt{\epsilon_{\ell} } (c^\dagger_j c_{j+1} + \text{h.c.}), \qquad j=1,...,L \label{eq:lind.imp.currentLind}
\end{eqnarray}

$\bullet$ Coherent Hamiltonian competing with the dissipative dynamics:
\begin{eqnarray}
\hat H = -\epsilon_H \sum_j (c^\dagger_j c_{j+1} + \text{h.c.}),
 \label{eq:lind.imp.H}
 \end{eqnarray}
describing the case in which despite the dissipation being the leading term, there are still some non-negligible coherent local dynamics within the fermionic system. 
We consider the simplest form of hopping terms, but small local coherent perturbations lead to the same conclusions.
 
$\bullet$ Additional set of decay dissipative channels:
\begin{eqnarray}
\hat \ell_{j,\rm decay} = \sqrt{\epsilon_{\rm decay}} \hat c_{j}, \qquad j=1,...,L
 \label{eq:lind.imp.decay}
\end{eqnarray}
corresponding to losses of particles in the optical lattice due to an extra channel, driving the fermionic system towards an empty vacuum state

\begin{figure}
 \includegraphics[scale=0.5]{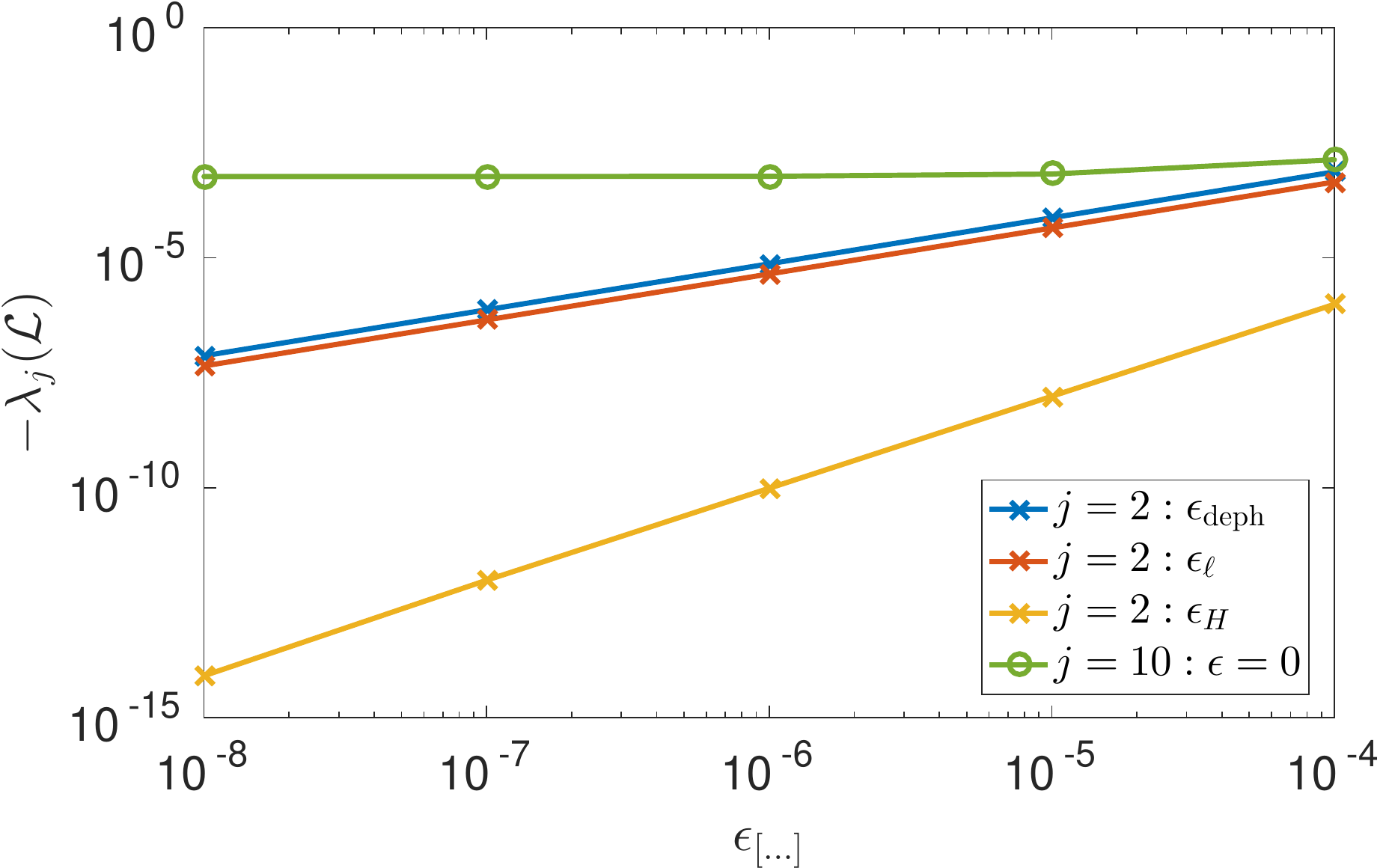}
 \caption{\textbf{Spectral properties of the Lindbladian with imperfections.} We consider the Lindbladian of the manuscript with parameters $A=B=\beta=t=1$ for a system with $L=9$ sites and analyse the effects of the imperfections described in Eqs.\eqref{eq:lind.imp.dephasing},\eqref{eq:lind.imp.currentLind},\eqref{eq:lind.imp.H} on the spectral properties of the Lindbladian. We see that in the presence of an imperfection the DDS is not completely degenerated anymore, with a splitting of the degeneracy proportional to the imperfection strength. We obtain that $\lambda_{1} = 0$ (by definition the Lindbladian has always a zero eigenvalue) while $\lambda_{j=2,3,...,9}<0$ and are approximately equal to each other. 
 We also show in the figure for clarity the eigenvalue $\lambda_{j=10}$ for $\epsilon = 0$,   corresponding to the Lindbladian gap in the unperturbed case, which is also approximately equal to the cases with imperfections in the considered range of $\epsilon$ strengths.
 }
 \label{fig:spectrumLindbladian_imperfections}
\end{figure}

Our results for the first three imperfections above are shown in Fig.\eqref{fig:spectrumLindbladian_imperfections}. We see that perturbation theory provides a qualitative picture: 
 in the regime of small perturbations (in units of the rate of the original Lindbladian, $\epsilon \ll 1$) while imperfections in the jump operators lead to a  
 linear splitting of the DDS, 
 $\lambda_2 \sim \epsilon$, a Hamiltonian perturbation leads to a quadratic dependence $\lambda_2 \sim \epsilon^2$.
Thus, as long as the perturbation is small compared to the unperturbed gap 
there is a time window between the system
entering the DDS and the system characteristics of the state being destroyed by the imperfections, which gives a possibility to effectively use these states for quantum information tasks.

The case of an additional set of decay dissipative channels follows in a similar form. In this case the steady state of the evolution is the vacuum state for $\epsilon_{\rm decay} > 0$. However, as above, if the perturbation is small there is a time window over which the effects on the DDS are negligible. One may obtain the characteristic time of the dissipative decay effects by the dynamics of the total number of particles $\hat N$ in the system, which in the Heisenberg picture is described by 
$\mathcal{L}^\dagger[\hat N] = - \epsilon_{\rm decay} \hat N$, i.e., $N(t) \sim e^{-\epsilon_{\rm decay} t}$. Thus the effects of particle losses in the system are relevant for times of the order $t \sim 1/\epsilon_{\rm decay}$, similarly to the other imperfections in the quantum jump operators considered above.

\section{Adiabatic evolution}
\label{sec.Adiabatic evolution}

In this Section we expand the discussion of the adiabatic evolution in the dissipative dynamics. 
Using the same protocol as in the main manuscript, we evolve the system from an initial state given by a superposition of the three charge density wave configurations, precisely, 
$|\psi(t=0)\rangle = \left( |\Psi_{a=0} \rangle + 
\sqrt{2} |\Psi_{a=1} \rangle  + 
\sqrt{3} |\Psi_{a=2} \rangle\right)/\sqrt{6} $ where $|\Psi_{a} \rangle $ are the Laughlin states for $\beta = 0$ (i.e., product charge density wave configurations).
We then evolve
this state with the Lindblad operators (Eq.\eqref{eq.lind.operators}) using a time-dependent $\beta$ parameter: $\beta(t) =\Delta \cdot t$ for $0 < t \leq 1/\Delta$
and $\beta(t) = 1$ for $t > 1/\Delta$, where $\Delta$ is the ramp velocity.

We analyse the purity $\gamma(t)$ of the quantum state during the dynamics, as well as its overlap onto the DDS. For the later we first describe the $3 \times 3$ density matrix 
$\hat \rho_{\rm DDS}$ representing the expectation value of the quantum state in the Laughlin dark states,
\begin{equation}
 (\hat \rho_{\rm DDS}(t))_{a,a'} \equiv 
 \langle \Psi_a(t) | \hat \rho(t) |\Psi_{a'}(t) \rangle 
\end{equation}
for $a,a'=0,1,2$, where $|\Psi_a(t) \rangle$ are the Laughlin dark states for the Lindblad parameters at time $t$. 
We study (i) the projection of the quantum state over the DDS, given by the diagonal terms of the matrix, $D_{\rm DDS}(t) = \sum_a \rho_{\rm DDS}(t)_{a,a'}$; (ii) how the coherence of the initial state evolve under the adiabatic evolution, quantified by
\begin{equation}
C(t) = \sqrt{\sum_{a \neq a'} \left(\frac{  \rho_{\rm DDS}(t)_{a,a'} - \rho_{\rm DDS}(0)_{a,a'}}{  \rho_{\rm DDS}(0)_{a,a'} } \right)^2},
\end{equation}
and (iii) the distinguishability of the full matrix with the initial state, quantified by the trace norm as follows,
\begin{equation}
 \mathcal{D}(t) = ||\hat \rho_{\rm DDS}(t)_{a,a'} - 
 \hat \rho_{\rm DDS}(0)_{a,a'}||_1
\end{equation}
Both coherence $C(t)$ as the distinguishability $\mathcal{D}(t)$ should be small if the evolved quantum state do not differ significantly from the initial state. 

\begin{figure}
\includegraphics[scale=0.5]{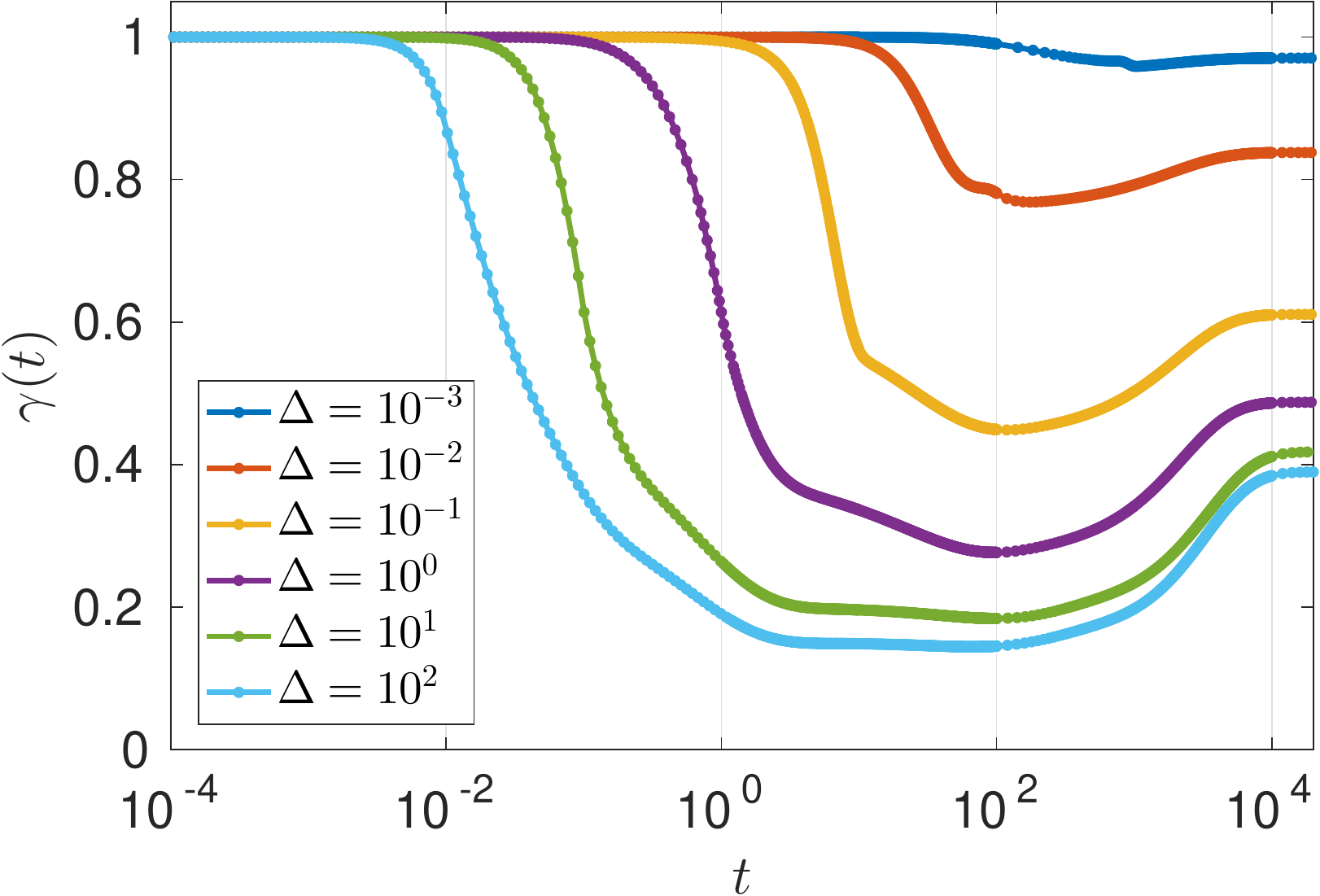}
\includegraphics[scale=0.5]{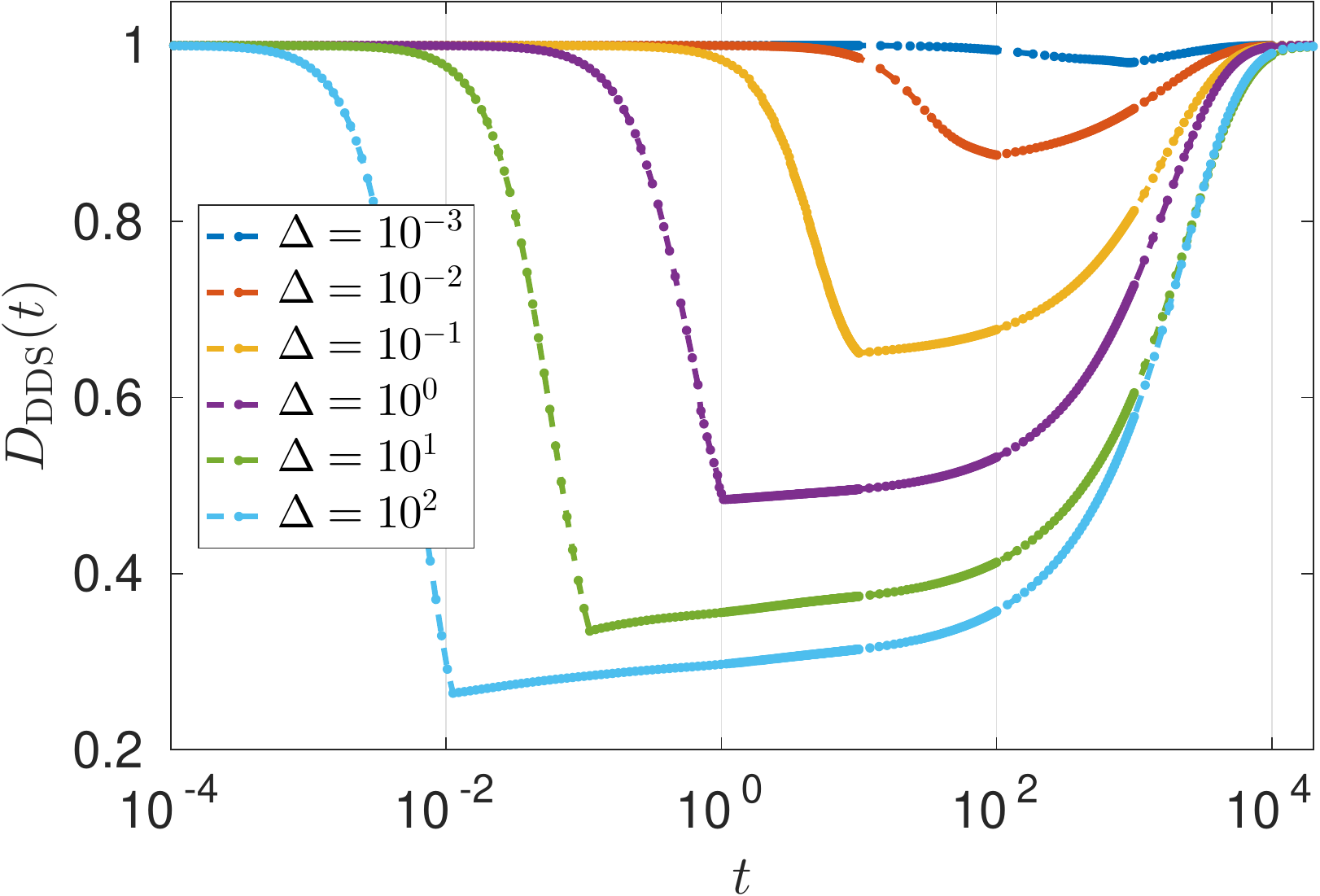}
\includegraphics[scale=0.5]{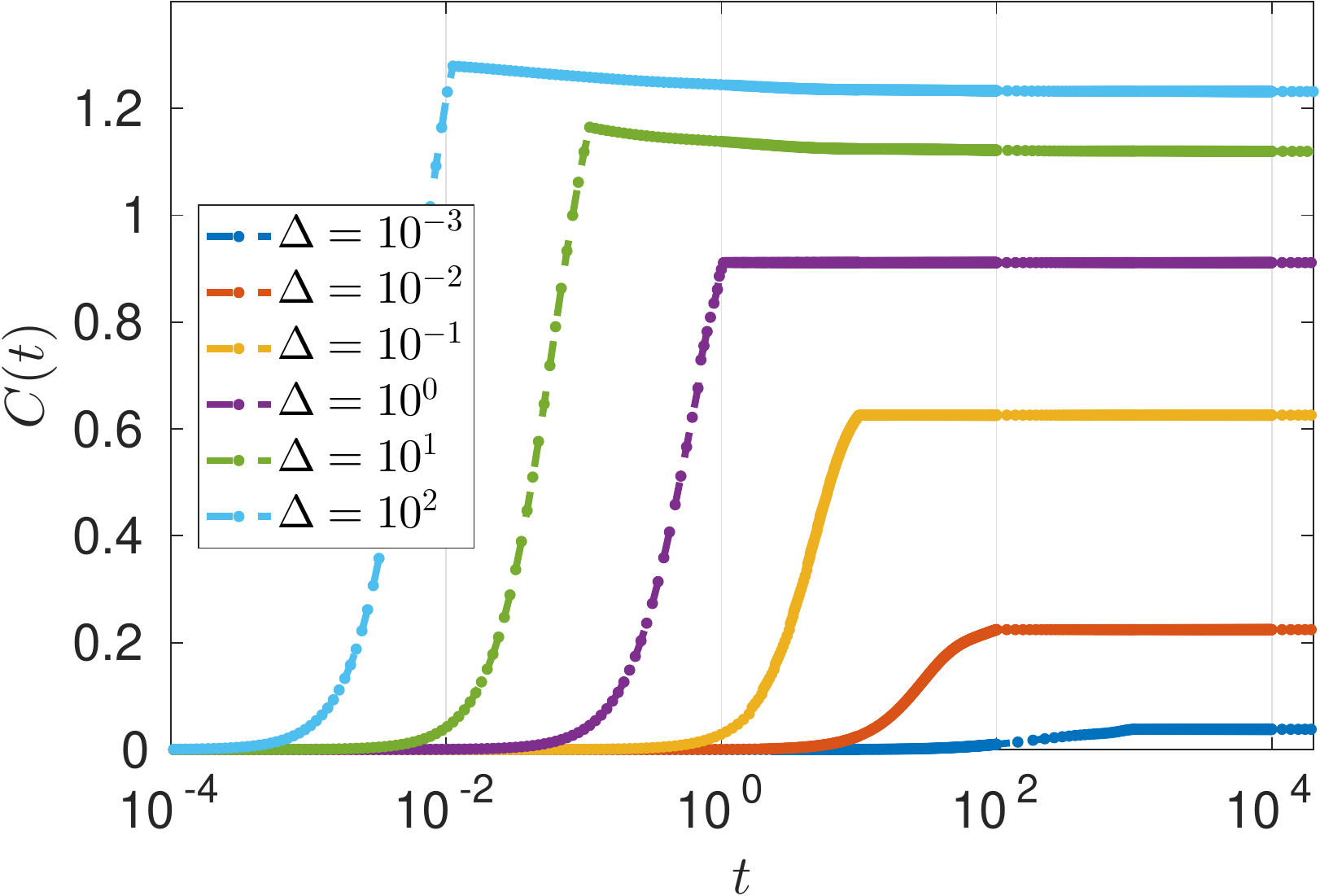}
\includegraphics[scale=0.5]{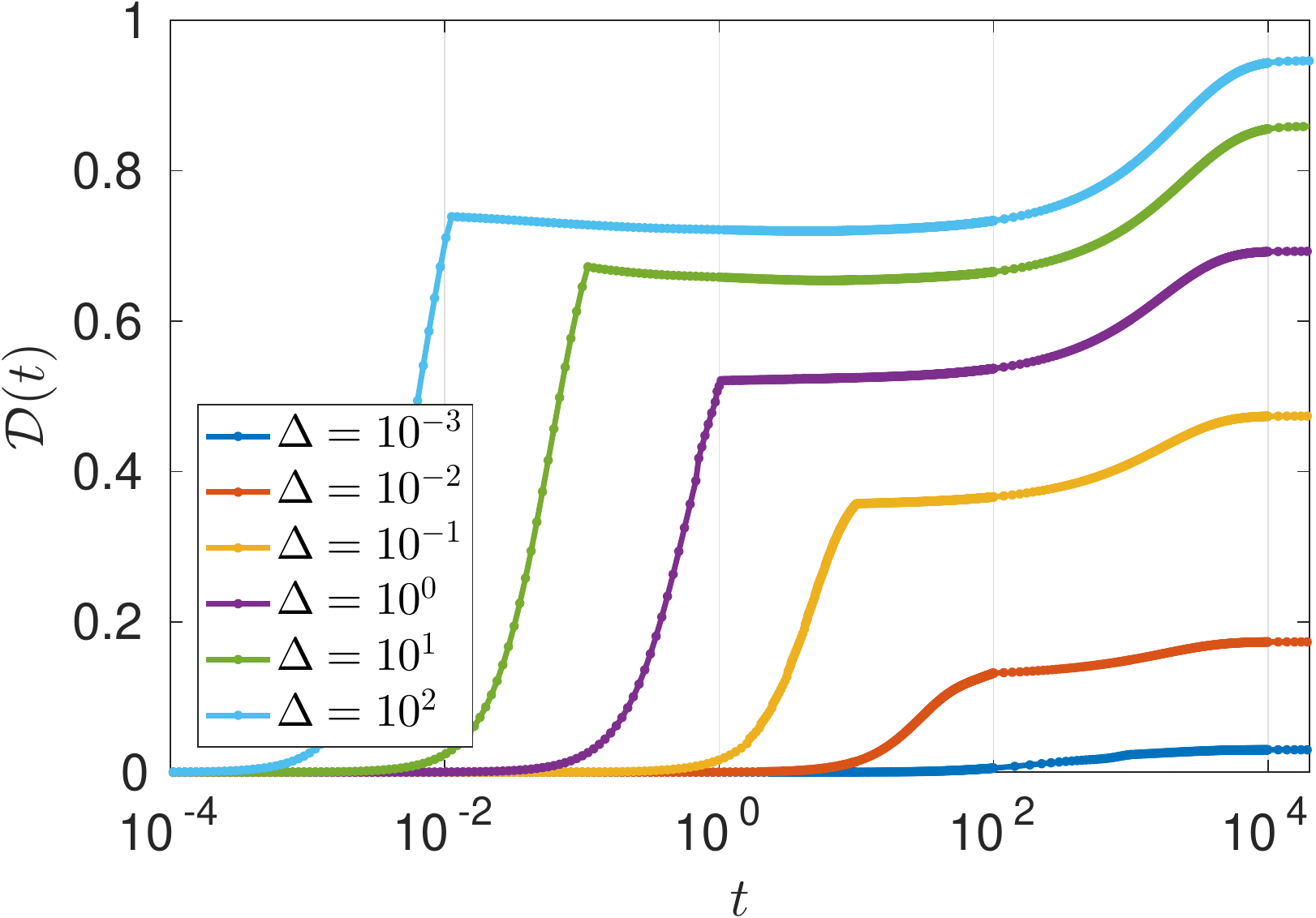}
\includegraphics[scale=0.5]{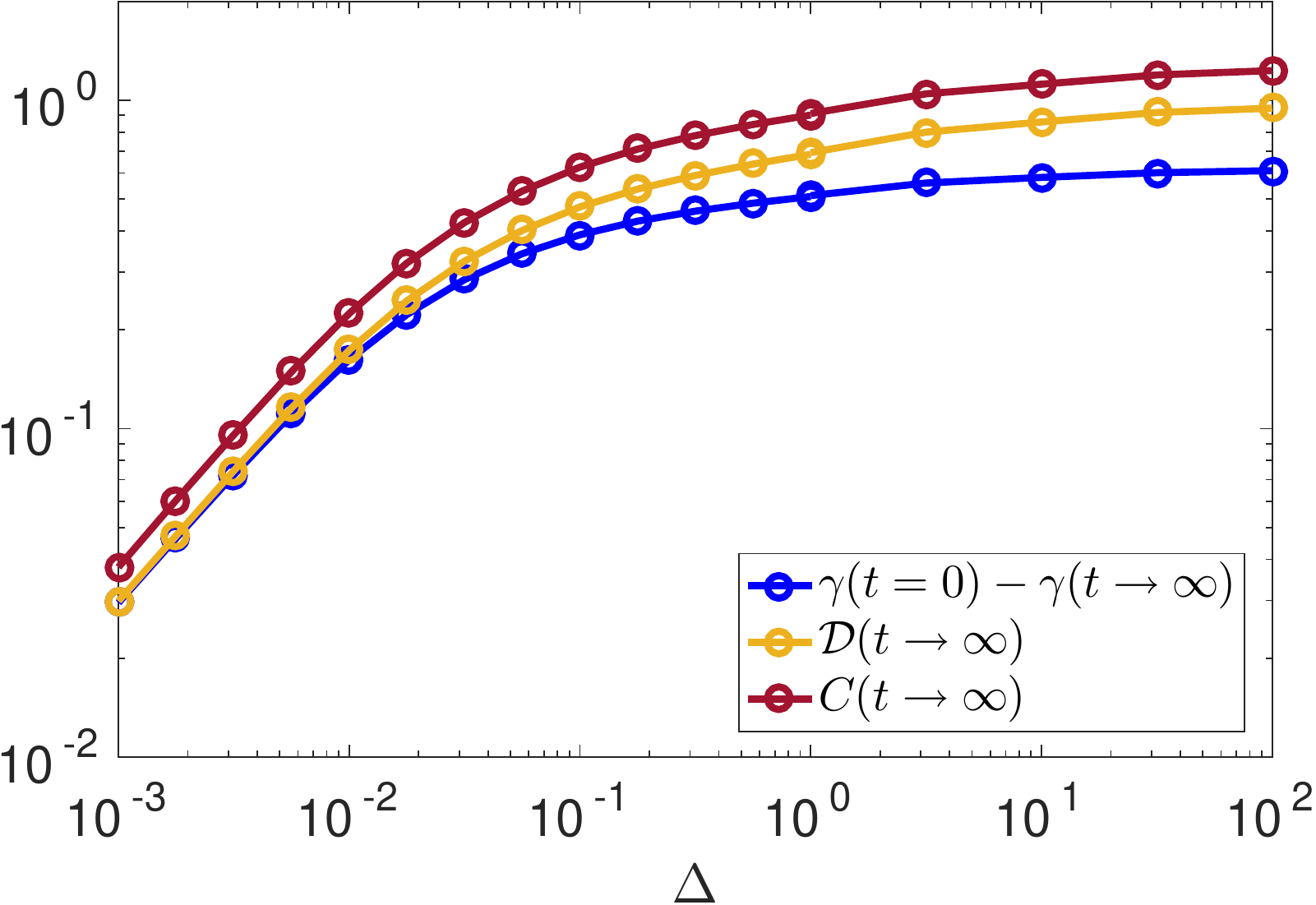}
 \caption{\textbf{Adiabatic evolution.} We consider a system with $L=9$ sites, $A=B=t=1$ and varying $\beta(t)$ adiabatically according to the ramp rate $\Delta$. The initial state of the system is given by the superposition $|\psi(t=0)\rangle = \left( |\Psi_{a=0} \rangle + 
\sqrt{2} |\Psi_{a=1} \rangle  + 
\sqrt{3} |\Psi_{a=2} \rangle\right)/\sqrt{6} $.
 We show the dynamics of the \textbf{(top-left)} purity, \textbf{(top-right)} $D_{\rm {DDS}}(t)$, 
 \textbf{(middle-left)} coherence $C(t)$ and 
 \textbf{(middle-right)} distinguishability $\mathcal{D}(t)$. In the \textbf{bottom} panel we show their steady state values.
 }
 \label{fig:adiabatic.ev.L9}
\end{figure}

 We show our results in Fig.\eqref{fig:adiabatic.ev.L9}.
 We see that for slow ramp rates the initial quantum state characteristics (purity and coherences) are preserved. We find in particular  that for very slow rates $\Delta \ll 1$ the steady state properties have  polynomial corrections compared to their initial conditions, i.e.,  $\gamma(t \rightarrow \infty) - \gamma(0)$, $C(t\rightarrow \infty)$ and  $\mathcal{D}(t\rightarrow \infty) \sim \Delta^{\rm cte}$.
   Interestingly also to notice that the coherence is approximately constant after an initial time $t \sim O(1/\Delta)$ (i.e., $\beta(t) \sim 1$), while it is the diagonal terms of the DDS subspace ($D_{\rm DDS}(t)$) that still shows non trivial dynamics after this initial transient time.

\section{Mapping Fractional Quantum Hall groundstate to one
dimensional model.}

In this section we revisit the exact mapping of the Laughlin state
\cite{Laughlin1981} into a one dimensional state\cite{Lee2004,Seidel2005}.
We will be interested in filling fractions $\nu<1.$ Recalling that
a 2DEG in a strong magnetic field displays Landau levels, we assume
that the relevant physics occurs in the lowest Landau level (LLL).
We place the system into a 2D torus
with linear sizes $L_{x}$ and $L_{y}$ and area $A=L_{x}L_{y}\sin\theta$,
defined by the region in the upper half complex plane enclosed by
the points $w=(0,L_{x},L_{y}\tau,L_{x}+L_{y}\tau)$. This torus is
characterized by the modular parameter $\tau=L_{y}/L_{x}e^{i\theta}=\tau_{1}+i\tau_{2}$,
($\rm{Im}(\tau)>0,\theta\in[0,\pi]).$ Following \cite{Haldane1985}
we introduce the translation operators $t(\mathbf{\mathbf{L}})=\exp(\mathbf{L}\cdot(\mathbf{\nabla}-ie\mathbf{A})-iL_{x}y+iL_{y}x)$
(here $\mathbf{L}=(L_{x},L_{y})$ are measured in units of the magnetic
length) which correspond to the usual translation operators in terms
of the canonical momentum, and an extra space dependent phase. The
single particle wavefunction satisfies the boundary conditions $t(\mathbf{L_{a}})\Psi=e^{i\phi_{a}}\Psi$,
with $\mathbf{L_{a}}$ a translation over the lattice vectors $\mathbf{L_{1}}=(L_{x},0)$
and $\mathbf{L_{2}}=L_{y}(\cos\theta,\sin\theta)$. Both conditions
can be satisfied if the flux over the torus $\frac{L_{x}L_{y}\sin\theta}{2\pi}=N_{\Phi}$
is integral. We parameterize the coordinates on the torus by $z=\tilde{z}/L_{x}$
with $\tilde{z}=L_{x}(x+y\tau),$ where $x\in[0,1]$ and $y\in[0,1]$.

The relation with the usual Cartesian coordinates is $x_{1}=L_{x}(x+\tau_{1}y)$ and $x_{2}=L_{y}\tau_{2}y$
The single particle wavefunction has the form $\Psi=e^{-\frac{1}{2}(\rm{Im}(\tau)L_{1}y)^{2}}f(z)$,
where $f(z)$ is an entire (holomorphic) function in the complex plane.
The we use units where $\sqrt{\hbar/eB}=1$. In the Landau gauge $\mathbf{A}=-By\hat{x}$, the boundary conditions read
\begin{eqnarray}\label{boundary_conds}
f(z+1)  =  f(z)e^{i\phi_{1}},\quad f(z+\tau) =  f(z)e^{i\phi_{2}}e^{-i\pi N_{\Phi}(2z+\tau)},
\end{eqnarray}
where the phases $\phi_{a}$ correspond to the fluxes piercing the torus in the
two orthogonal directions $a=1,2.$ From these relations it follows that $\int dz\frac{d}{dz}\ln(f(z))=2\pi iN_{\Phi}$,
which implies that the function $f(z)$ has $N_{\Phi}$ zeroes. The single particle wavefunction
that satisfies the boundary conditions (\ref{boundary_conds}) and has $N_{\Phi}$
zeroes is given by the generalized theta function
\begin{eqnarray}\label{single_part_LLL}
\phi_{n}(z;\tau,\phi_{1},\phi_{2}) & = & e^{-\frac{1}{2}({\rm Im}(\tau)L_{x}y)^{2}}\vartheta\left(z-z_n\left|\frac{\tau}{N_{\Phi}}\right.\right)e^{i\phi_{1}\left(z-z_n\right)}
\quad\mbox{with}\quad\vartheta(z|\tau)  =  \sum_{m=-\infty}^{\infty}\left(e^{i\pi\tau}\right)^{m^{2}}e^{2\pi imz}
\end{eqnarray}
and $z_n=\frac{2\pi n+\phi_{2}-\tau\phi_{1}}{2\pi N_{s}}$.
This corresponds to a normalizable wavefunction for ${\rm Im}(\tau)>0$.
The zeroes of $\varphi_{n}(z;\tau,\phi_{1},\phi_{2})$ are located
at $z=z_n+\frac{1}{2}+m+\left(\frac{1}{2}+n\right)\frac{\tau}{N_{\Phi}}$.

As shown in Ref \onlinecite{Trugman1985}, the Laughlin
state at filling $\nu=1/3$ is the zero energy exact ground state
of the Landau problem with the interaction $\mathcal{H}=V_{0}\int d\bm{r}|\nabla \rho(\bm{r})|^2$, where $\rho(\bm{r})=\psi^\dagger(\bm{r})\psi(\bm{r})$
and  $\bm{r}=(x_{1},x_{2})$. The projection
of the electron operator into the first Landau level is $\psi=\sum_{n}\varphi_{n}(\mathbf{r})c_{n},$
where $c_{n}$ destroys a state at occupation $n$. The interaction Hamiltonian
projected onto the first Landau level becomes
\begin{equation}
\mathcal{H}=\frac{32N_{\Phi}V_{0}}{|\tau|^{3}L_{1}^{2}}\sum_{j}Q_{j}^{\dagger}Q_{j}\quad\mbox{with}\quad Q_{j}^{\dagger}=\sum_{j}^{N_{\Phi}}\sum_{k=-\infty}^{\infty}\left({j}-N_\Phi k\right)e^{-\frac{2\pi i}{N_\Phi\tau}\left(j-kN_\Phi\right)^{2}}c_{j+k}^{\dagger}c_{j-k}^{\dagger}.\label{eq:destruction_laughlin}
\end{equation}
In this sum the pair of numbers $(j,k)$ are all integers or all half integers and satisfy $0 < j < N_\Phi , 0 < k, l, < N_\Phi/2$. Separating both cases, and defining
$\kappa^2= \frac{2\pi}{N_\Phi}\frac{L_x}{L_y}$ gives the operators $\ell_{s,n}$ (Eq. 5) in the main text.

\section{Physical Realization}

We consider a one dimensional optical lattice (system) immersed in
condensate that acts as a bath to the system, providing dissipation.
Each site in the optical lattice consists of a potential-well accommodating
two single particle levels denoted by $c$ and $f$
\begin{equation}
H_{{\rm sys}}=-\sum_{i,\sigma}^{N}(J_{\sigma}a_{i,\sigma}^{\dagger}a_{i+1,\sigma}+\text{h.c})-U\sum_{i}n_{i,1}n_{i+1,1}+\sum_{i,\sigma}E_{\sigma}n_{i\sigma},\label{eq:Hamiltonian}
\end{equation}
Here $a_{i,\sigma}$ is a fermionic annihilation operator at site $i=1\dots N_\Phi$ and level
$\sigma=\{0,1\}$. The relation with the main text operators is $a_{i,0}=c_i$ and $a_{i,1}=f_i$. The Hamiltonian includes an attractive ($U>0$) interaction between neighbour particles in the level $\sigma=1$.
The operator $n_{i\sigma}=a_{i\sigma}^{\dagger}a_{i\sigma}$ measures the occupation at the site $i$ and level $\sigma$. We assume that the number of particles in 
the system is given by $N_{e}=N_\Phi/3$.

To capture the essential physics generated by the interaction, we
first study the two particle problem. Defining the two particle state
$|\sigma\sigma'\rangle=\sum_{i,j}\chi_{ij}^{\sigma\sigma'}a_{i,\sigma}^{\dagger}a_{j,\sigma'}^{\dagger}|0\rangle$,
with $|0\rangle$ the state with no particles (vacuum), the Schr\"{o}dinger
equation for the wavefunction $\chi_{ij}^{s}=\frac{1}{\sqrt{2}}(\chi_{ij}^{11}-\chi_{ji}^{11})$ reads 
$-J_{1}(\Delta_{i}+\Delta_{j})\chi_{ij}^s-U(\delta_{i,j+1}+\delta_{i+1,j})\chi^s_{ij}=(E-2E_{1}+4J_{1})\chi_{ij}^{s},$
 with $\Delta_{i}\chi_{ij}=\chi_{i+1,j}-2\chi_{ij}+\chi_{i-1,j}$,
the discrete Laplace operator. Introducing the central and relative coordinates $R=a(j+l)/2$ and $r=a(j-l)$
the wavefunction can be written as $\chi_{jl}=e^{iRK}\chi(K)_{r}=e^{iRK}\sum_{q}e^{irq}\tilde{\chi}_{q}^{s}$
where we have introduced the total and relative momentum $K=k_{1}+k_{2}$
and $2q=k_{1}-k_{2}$. The Schr\"odinger equation for $\chi_{ij}^s$ becomes 
\begin{equation}
\tilde{\chi}^s_{q}=\frac{2}{N_\Phi}\frac{U\sin q\bar{\chi}}{E-2E_{1}+4J_{1}\cos\frac{K}{2}\cos q}\rightarrow E_{d}(K)=2E_{1}-U-4\frac{J_1^2}{U}\cos^2 \frac{K}{2}\label{eq:Consistent}
\end{equation}
 where $\bar{\chi}=\sum_{q}2\sin q\tilde{\chi}_{q}^{s}$. For fixed center of mass momentum, the bound state energy $E_{d}(K)$
is found by solving self-consistently Eq. (\ref{eq:Consistent}).
We consider the regime $J_{0}\sim0$, along with $|U|\gg E_{1}$ and $\frac{J_{1}}{E_1}\ll\frac{J_{1}}{|U|}\ll1.$
In this case, the bound state energy $E_d(k\sim 0)$ is far below the bottom of the $(1,0)$-pair band, but still above the $(0,0)$ two-particle band. 
The amplitude for tunneling between two $0$ states is taken to be negligible compared to all other energy scales ($J_{0}\sim0$).
This implies a flat band for the $(0,0)$-pair. The two-particle energies $E_{\sigma\sigma'}$ of the continuous Bloch bands are 
\begin{equation}
E_{\sigma\sigma'}(K,q)=(\sigma+\sigma')E_{1}-2(J_{\sigma}+J_{\sigma'})\cos\frac{Ka}{2}\cos qa+2(J_{\sigma}-J_{\sigma'})\sin\frac{Ka}{2}\sin qa,
\end{equation}
A doublon state of definite momentum is created by the combination 
$d_{K}^{\dagger} =\sum_{R}e^{iKR}\sum_\ell e^{-\ell/\xi(K)}f_{R+\frac{\ell}{2}}^{\dagger}f_{R-\frac{\ell}{2}}^{\dagger},$ with
$\xi^{-1}(K)=\ln\left(\frac{J_1}{|U|}\cos \frac{K}{2}\right).$ The state $|d_{K}\rangle\equiv d_{K}^{\dagger}|0\rangle$ is normalized
as $\langle d_{K'}|d_{K}\rangle=\delta_{K'K}$, with $|0\rangle$
a state with no particles.

\section{Laser Driving and coupling to a bath}

We are interested in the dynamics generated between the low lying doublon states and the $(0,0)$ lower energy band. We can induce Raman transitions between states
in these bands using an external driving laser with Raman detuning $\Delta=2E_{1}-U-\omega$. The interaction
between the (classical) radiation of amplitude $\Omega$ and frequency $\omega$, and the system is 
$H_{\rm rad} = \Omega \cos(\omega t)\sum_i f^\dagger_{i}(c_{i}+\alpha(c_{i-1}+c_{i+1}))+\text{h.c}$ with $\alpha=\frac{A_1}{A_0}\ll 1$. 
The amplitudes $A_m$ decay fast with $m$ as they represent the matrix element of the different Wannier functions and the laser radiation profile.

We couple our system to a 3-dimensional (3D) Bose Einstein Condensate (BEC).
This coupling is realized by immersing the system in the BEC which acts as a dissipative, memoryless bath (valid when the spectral function of the bath is almost constant around the
frequency of the laser $\omega$). The bath is very efficient to de-excite the system and does not induce excitations to higher energy bands as
its temperature $T$ satisfy $T\ll\omega$ so thermal excitations in the bath cannot induce excitations in the system.
This type of bath has been used to obtain superconducting states realizing Majorana zero modes \cite{Diehl2011}. The quasiparticle excitations
of this bath are described by the Hamiltonian $H_{{\rm bath}}=\sum_{k}E_{k}b_{\mathbf{k}}^{\dagger}b_{\mathbf{k}}$,
where the bogoliubov quasiparticles $b_{\mathbf{k}}$ have mass $m_{b}$
and propagate with velocity $c_{b}$. These bosonic quasiparticles
have energy $E_{k}=k(c_{b}^{2}+k^{2}/4m_{b}^{2})^{1/2}$,with $k=\sqrt{\mathbf{k}^{2}}=(k_{x}^{2}+k_{y}^{2}+k_{z}^{2})^{1/2}$
the magnitude of the three dimensional momentum of the Bogoliubov
quasiparticles. Here $b_{\mathbf{k}}$ $(b_{\mathbf{k}}^{\dagger})$
destroys (creates) a bosonic bogoliubov excitation of momentum $\mathbf{k}$. 
We work in the regime where the excitation induced by the laser is
relaxed by the bath immediately, such that states with multiple excitations
are not created. This approximation
corresponds to the limit of weakly far detuned laser $E_{d}\gg|\Delta|\gg|g|,|\Omega|$.

The interaction between the system and the condensate is described
by the density interaction $H_{{\rm bath/sys}}=g\int d\mathbf{r}\delta\rho_{s}(\mathbf{r})\delta\rho_{{\rm bath}}(\mathbf{r})$
\cite{Diehl2008}, where $g$ is the strength of the system/bath coupling, which is assumed to be small. The density $\delta\rho_s(\mathbf{r})$
is the density of fermions at the point $\mathbf{r}$ in 3D, so it involves the Wannier wavefunctions of the fermions. The density $\delta\rho_{\rm bath}$
corresponds to the phonon waves around the equilibrium BEC density induced by the interaction with the fermions in the optical lattice.\cite{Diehl2008} 
We move to a rotating frame of reference (the rotating wave approximation,
RWA), where the problem, in the reduced Hilbert space which consists
of the two lowest bands and the bath, looks static. This is achieved
by introducing the time dependent unitary transformation $\mathcal{U}_{t}=\exp\left(iH_{{\rm sys}}t\right).$
The penalty is that terms involving higher bands of the extended Hilbert
space are fast oscillating. By coarse graining in time we are allowed
to get rid of these terms, which amounts to projecting out the higher
bands. The radiation Hamiltonian is modified by the rotating wave approximation
accordingly. In this rotating basis,
the interaction Hamiltonian with the bath $H_{{\rm bath/sys}}$ 
is also modified as $H'_{{\rm int}}=\mathcal{U}_{t}H_{{\rm int}}\mathcal{U}_{t}^{\dagger}$.
Retaining only the slow oscillating terms (with frequency $\sim\omega$)
in the system-bath interaction, amounts to consider just the transition between the doublon and the $(0,0)$ band. Each
of these transitions is accompanied by the creation (or destruction) of a phonon excitation in the bath.

Tracing over the bath, using the Born-Markov approximation, results in a quantum master equation
for the effective density matrix in the reduced Hilbert space involving the bottom and
the doublon bands. For small amplitude of the external drive $\Omega$, such that
$|\Delta|\gg|g|,|\Omega|$, no more than one doublon is excited
at any given time. We may then trace out the doublon band, and obtain
a closed dissipative equation of motion within the Hilbert space of
the lowest band. After this adiabatic elimination of the doublon band, we find
the quantum master equation 
$ \dot{\rho}=\mathcal{L}(\rho)=\sum_{i=1}^{N_\Phi} \gamma \left(\tilde{\ell}_i\rho \tilde{\ell}_i^{\dagger}-\frac{1}{2}\{\tilde{\ell}_i^{\dagger} \tilde{\ell}_i, \rho\}\right),$
with $\gamma=\frac{\Omega^2 g^2}{\Delta^2} \Gamma_0$ and $\Gamma_0$ parameterizing details of the bath/system coupling.
The quantum jump operator is in turn
$\tilde{\ell}_{i}= R^\dagger_{i}\mathcal{Q}_{i}$ with $R^\dagger_{i}= c_{i}^\dagger c_{i+1}^\dagger + t(c^\dagger_{i}c^\dagger_{i+2}+c^\dagger_{i+1}c^\dagger_{i+3}) $, and
$\mathcal{Q}_{i}= \ell_{1,i}+A(\ell_{0,i-1}+\ell_{0,i+1})+B(\ell_{1,i+1}+\ell_{1,i-1})$.

%


\end{document}